\documentclass[a4paper,UKenglish,cleveref, autoref, thm-restate]{lipics-v2021}

\usepackage{amsmath,amsfonts}
\usepackage{algorithmic}
\usepackage{graphicx}
\usepackage{textcomp}
\usepackage{xcolor}
\usepackage{xspace}
\usepackage{booktabs}
\usepackage{url}
\usepackage{hyperref}
\usepackage{makecell}
\usepackage{subcaption}
\usepackage{wrapfig}
\usepackage{caption}
\usepackage{cleveref}
\usepackage{siunitx}

\long\def\todo#1 { {\bf TODO:} [{\color{gray} #1}] }

\newcommand{\eg}{{\em e.g.}\xspace}
\newcommand{\ie}{{\em i.e.}\xspace}

\newcommand{\hash}[0]{\mathrm{H}\xspace}
\newcommand{\group}[0]{\ensuremath{\mathbb{G}}\xspace}
\newcommand{\ZZ}{\ensuremath{\mathbb{Z}}}

\newcommand{\sk}{\ensuremath{\mathrm{sk}}\xspace} 

\newcommand{\cpk}[1]{\ensuremath{\mathrm{pk}^{#1}_{\mathrm{smc}\xspace}}} 
\newcommand{\csk}[1]{\ensuremath{\mathrm{sk}^{#1}_{\mathrm{smc}\xspace}}} 

\newcommand{\tx}{\ensuremath{\mathrm{tx}}\xspace}
\newcommand{\txw}{\ensuremath{\mathrm{tx_w}}\xspace}
\newcommand{\txwp}{\ensuremath{\mathrm{tx_w'}}\xspace}

\newcommand{\enc}[1]{\ensuremath{\mathrm{enc}_{#1}}}
\newcommand{\dec}[1]{\ensuremath{\mathrm{dec}_{#1}}}
\newcommand{\sign}[1]{\ensuremath{\mathrm{sig}_{#1}}}

\newcommand{\System}{F3B\xspace}
\newcommand{\system}{F3B\xspace}
\newcommand{\SMCa}{SMC\xspace}
\newcommand{\smca}{SMC\xspace}

\newcommand{\smc}{secret-man\-age\-ment committee\xspace}

\newcommand{\smcs}{secret-man\-age\-ment committees\xspace}
\newcommand{\bc}{block\-chain\xspace}
\newcommand{\ub}{underlying block\-chain\xspace}
\newcommand{\cg}{consensus group\xspace}
\newcommand{\cns}{consensus nodes\xspace}
\newcommand{\cn}{consensus node\xspace}

\long\def\comment#1{}

\long\def\com#1{}

\hideLIPIcs  

\title{F3B: A Low-Overhead Blockchain Architecture with Per-Transaction Front-Running Protection} 

\titlerunning{Flash Freezing Flash Boys(F3B)} 

\author{Haoqian Zhang}{École Polytechnique Fédérale de Lausanne, Switzerland}{haoqian.zhang@epfl.ch}{}{}

\author{Louis-Henri Merino}{École Polytechnique Fédérale de Lausanne, Switzerland}{louis-henri.merino@epfl.ch}{}{}

\author{Ziyan Qu}{École Polytechnique Fédérale de Lausanne, Switzerland}{ziyan.qu@epfl.ch}{}{}

\author{Mahsa Bastankhah}{École Polytechnique Fédérale de Lausanne, Switzerland}{mahsa.bastan76@gmail.com}{}{}

\author{Vero Estrada-Galiñanes}{École Polytechnique Fédérale de Lausanne, Switzerland}{vero.estrada@epfl.ch}{}{}

\author{Bryan Ford}{École Polytechnique Fédérale de Lausanne, Switzerland}{bryan.ford@epfl.ch}{}{}

\authorrunning{H. Zhang et al.} 

\Copyright{Haoqian Zhang, Louis-Henri Merino, Ziyan Qu,
Mahsa Bastankhah, Vero Estrada-Galiñanes, and Bryan Ford} 

\ccsdesc[500]{Security and privacy~Distributed systems security}

\keywords{Blockchain, DeFi, Front-running Mitigation} 

\category{} 

\nolinenumbers 

\EventEditors{Joseph Bonneau and S. Matthew Weinberg}
\EventNoEds{2}
\EventLongTitle{5th Conference on Advances in Financial Technologies (AFT 2023)}
\EventShortTitle{AFT 2023}
\EventAcronym{AFT}
\EventYear{2023}
\EventDate{October 23-25, 2023}
\EventLocation{Princeton, NJ, USA}
\EventLogo{}
\SeriesVolume{282}
\ArticleNo{3}

\begin{document}

\maketitle

\begin{abstract}
Front-running attacks, which benefit from advanced knowledge 
of pending transactions, have proliferated in the blockchain space since the emergence of decentralized finance.
Front-running causes devastating losses to honest 
participants and continues to endanger the fairness of the ecosystem.
We present Flash Freezing Flash Boys (\system), 
a blockchain architecture that addresses front-running attacks by using 
threshold cryptography.
In \system, a user generates a symmetric key to encrypt their transaction,
and once the underlying consensus layer has finalized the transaction, 
a decentralized \smc reveals this key.
\System mitigates front-running attacks
because, before the consensus group finalizes it, an adversary can no longer read the content of a transaction, thus preventing the adversary from benefiting from advanced knowledge of pending transactions.
Unlike other mitigation systems, 
\System properly ensures that all unfinalized transactions,
even with significant delays, 
remain private
by adopting per-transaction protection.
Furthermore, \System addresses front-running at the execution layer;
thus, our solution is agnostic to the underlying consensus algorithm and 
compatible with existing smart contracts.
We evaluated \system on Ethereum 
with a modified execution layer
and found only a negligible (0.026\%) increase in transaction latency, 
specifically due to running threshold decryption 
with a 128-member secret-management committee 
after a transaction is finalized; 
this indicates that \system is both practical and low-cost.
\end{abstract}

\section{Introduction}
Front-running is the practice of
benefiting from the advanced knowledge of pending
transactions~\cite{eskandari2019sok,bernhardt2008front,NasdaqFrontRunning}.
Although benefiting some entities involved, this practice puts
others at a significant financial disadvantage, 
making this behavior illegal in traditional markets 
with established securities regulations~\cite{eskandari2019sok}.

However, the open and pseudonymous nature of blockchain transactions
and the difficulties of pursuing miscreants across numerous 
jurisdictions have made front-running
attractive, particularly in decentralized 
finance (DeFi) \cite{lewis2014flash,eskandari2019sok,daian2020flash}.
Front-running actors in the blockchain space 
can read the contents of pending transactions
and benefit from them by, \eg, creating their own transactions
and positioning them according to the 
target transaction~\cite{baum2021sok,daian2020flash,eskandari2019sok}.
%

Front-running negatively impacts honest DeFi actors
and endangers the fairness of this multi-billion market~\cite{defipulse}.
One estimate suggests that front-running attacks amount
to \$280 million in losses for DeFi actors each 
month~\cite{losedefi}. 
Front-running also threatens the underlying consensus 
layer's security by incentivizing unnecessary
forks~\cite{dapp2020uniswap,daian2020flash}.

\begin{figure}[t]
    \centering
    \includegraphics[scale=0.253]{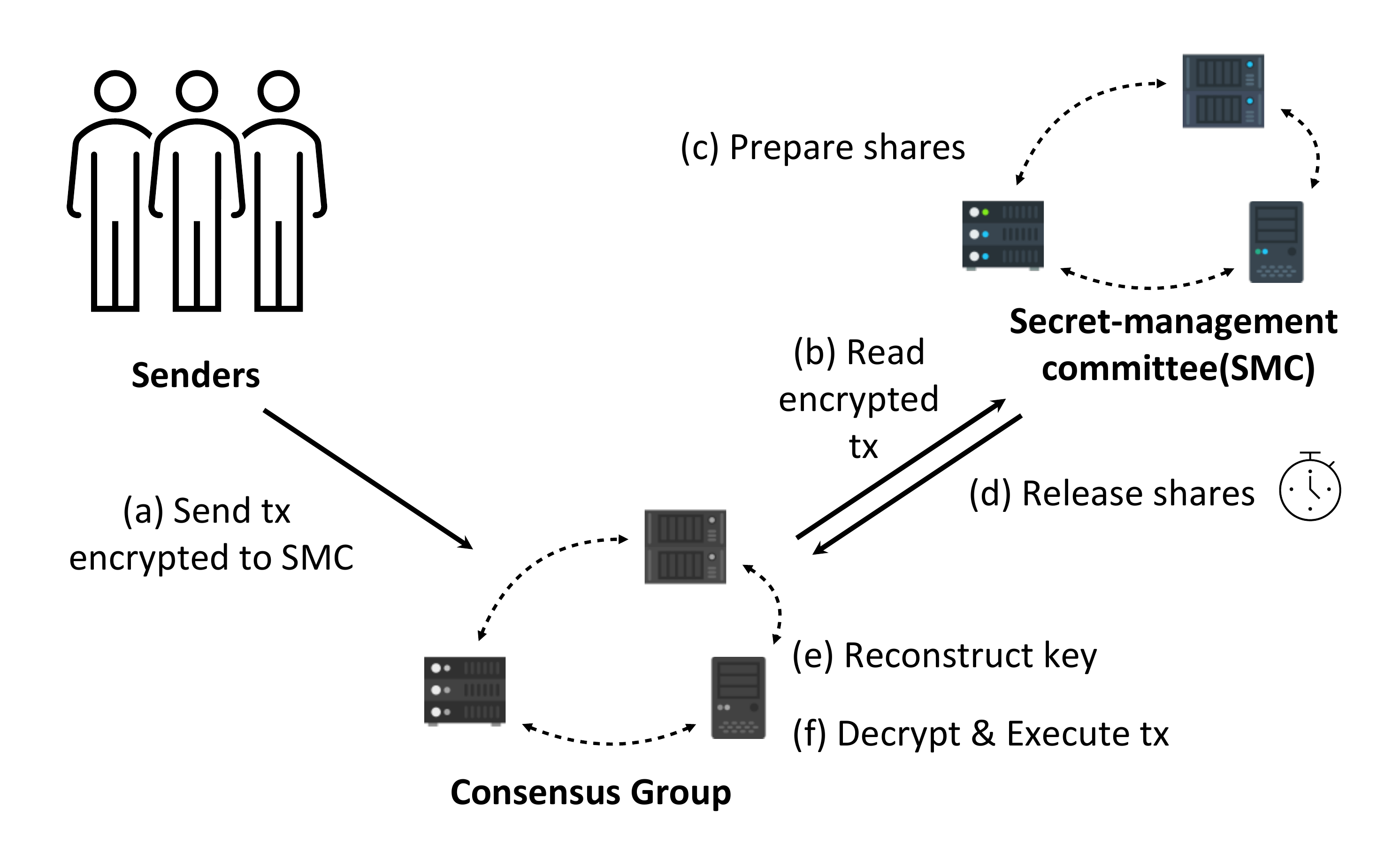}
    \caption{\system architecture. Senders publish encrypted transactions to the consensus group.
    The \smc releases the decryption shares
    once the transactions are no longer pending.
    Finally, the \cg reconstruct the key and 
    decrypt and execute the transaction.
    The \smc and the consensus group
    can consist of the same set of servers.
    For clarity in this paper, 
    we logically separate them into two different entities. 
    }
    \label{Fig:architecture}
\end{figure}

Despite work addressing front-running, 
several unmet challenges exist, 
such as high latency, 
being restricted to a specific environment, 
or raising security concerns.
Namecoin, an early example of mitigating front-running attacks 
by having users send a commit and later a reveal transaction,
requires two rounds of communication 
with the \ub~\cite{kalodner2015empirical}.
%
Submarine further improves Namecoin's design 
by hiding the addresses of smart contracts involved,
but it induces three rounds of communication 
to the \ub~\cite{libsubmarine,kalodner2015empirical}.
Both approaches induce high latency.
Other works have taken a different approach
to mitigate front-running attacks by tailoring their solution
to a specific application or consensus 
algorithm~\cite{ciampi2021fairmm,noah2021secure,stathakopoulou2021adding,bebel2022ferveo,ankalkoti2017relative,asayag2018helix,malkhi2022maximal,flashbotsprotection}.

A promising approach is to use threshold encryption,
where clients encrypt their transactions to 
prevent malicious actors from understanding
those transactions, as presented in 
Fairblock~\cite{momeni2022fairblock} 
and Shutter~\cite{shutter,shutterL1}.
However, these schemes require clients to choose a future block
to derive the encryption key, 
which raises security concerns.
Suppose a transaction failed to be finalized in the client-chosen block
due to, for example, a crypto mania that overwhelms the blockchain network~\cite{CryptoKitties}
or a deliberate denial of service attack~\cite{eskandari2019sok}.
In this case, the transaction is undesirably revealed 
(see \Cref{sec:StrawmanIII} for details).

We present Flash Freezing Flash 
Boys\footnote{
	The name \emph{Flash Boys} comes from a popular book 
	revealing this aggressive market-exploiting 
	strategy on Wall Street in 2014~\cite{lewis2014flash}.
}
(\system), a novel \bc architecture with front-running protection
that has a low latency overhead and is compatible with existing
consensus algorithms and smart contract implementations.
Like Fairblock~\cite{momeni2022fairblock} and Shutter~\cite{shutter,shutterL1},
\System addresses front-running 
by adopting threshold encryption, but it
accomplishes this on a \textbf{per-transaction} basis 
rather than a per-block basis.
Rather than selecting an encryption key 
linked to a future block, 
clients generate an encryption key for each transaction. This ensures that a transaction remains confidential
until the block containing the transaction has received enough confirmations.

As described in \Cref{Fig:architecture},
\System's architecture consists of the following steps: 
(a) A client encrypts their transaction to a \smc (\smca) and
sends their encrypted transaction to the \cg
that operates the underlying blockchain.
(b) The \smca reads the encrypted transaction from the \ub.
(c) The \smca prepares the decryption shares for the \cg.
(d) The \smca releases the decryption shares to the \cg
once the \ub has finalized the transaction.
(e) The \cg reconstructs the key.
(f) The \cg decrypts and executes the transaction.
Once the \smca begins to release the decryption shares,
malicious actors cannot launch
a front-running attack 
because the transaction is already irreversibly ordered on the blockchain.
Although adversaries may attempt to run 
\emph{speculative} front-running attacks, 
where they guess the contents of a transaction 
on metadata information like the sender's address, 
these attacks are more likely to fail 
and can prove to be unprofitable~\cite{baum2021sok}. 
Nonetheless, we discuss mitigation solutions for these attacks in \Cref{sec:metadata}.

\system addresses two key practical challenges:
(a) mitigating spamming of inexecutable encrypted transactions onto the \ub,
and (b) limiting latency overhead.
To mitigate spamming,
we introduce a deposit-refund storage fee for storing encrypted transactions,
along with the standard execution fee (\eg, gas in Ethereum).
To limit the latency overhead, 
users write only data onto the \ub once 
to achieve front-running protection.

We propose two cryptographic threshold schemes
that can plug into \System:
TDH2\cite{shoup1998securing} and PVSS~\cite{schoenmakers1999simple}.
TDH2 enables clients to encrypt their transactions 
under the same public key of a \smc
which is only changeable by time-consuming DKG or resharing protocols.
On the other hand,
PVSS empowers clients to adopt a different \smc for each transaction
but at the cost of the additional preprocessing time for preparing the shares
for each transaction.

We implemented a prototype of \system
with post-Merge\footnote{The Merge refers to the merge executed on September 15th, 2022, to complete Ethereum's transition to proof-of-stake consensus.} Ethereum~\cite{ethereumMerge} as the \ub
and Dela~\cite{Dela} as the \smc.
We measure the latency overhead 
by comparing the time 
it takes to decrypt and execute a transaction 
with the time it takes just to execute the transaction.
Our analysis shows that, with a committee size of 128,
the latency overhead is 0.026\% and 0.027\% for Ethereum 
under the TDH2 and PVSS respectively;
In comparison, Submarine,
which also offers per-transaction protection and hides the address of smart contracts as \system, exhibits a 200\% latency overhead, as
it requires three rounds of communication with the \ub~\cite{libsubmarine,breidenbach2018enter}. 
For part of our prototype, we modified Ethereum's execution layer
by adding a new transaction type featuring encryption and delayed execution.
By only modifying the execution layer,
we can (a) provide compatibility with various consensus algorithms 
embedded in Ethereum's consensus layer,
including Proof-of-Work (PoW),
Proof-of-Authority (PoA) and the
recently added, Proof-of-Stake (PoS)
and (b) protect existing smart contracts 
without requiring any code modifications.

In this paper, our key contributions are as follows:
\begin{enumerate}
    \item The design of a blockchain architecture with front-running protection that
    uses threshold encryption on a per-transaction basis, enabling
    confidentiality for all pending transactions, 
    even if transactions are delayed,
    while achieving low overhead.
    \item The design of two protocols based on TDH2 and PVSS for \system
    satisfies various demands and user scenarios.
    \item A prototype that, on Ethereum's execution layer, demonstrates
    \system's ability to be agnostic to (a) the underlying consensus algorithm 
    and (b) to smart contract implementations while achieving low-latency overhead.
    \item A systematic evaluation of \system on post-Merge Ethereum
    by looking at transaction latency, throughput, and reconfiguration costs.

\end{enumerate}

\section{Background}

In this section, we present a brief background on blockchain and smart contracts,
and we introduce front-running attacks and mitigation strategies.

\subsection{Blockchain \& Transaction Ordering}
A blockchain is an immutable append-only 
ledger of ordered transactions~\cite{nakamoto2008bitcoin}.
However, transactions go through a series of stages before they are
finalized---irreversibly ordered---on the blockchain.
After a sender creates a transaction, they need to propagate the transaction
among the consensus nodes that then place the transaction in a pool of pending transactions,
most commonly known as mempool.
Notably, these transactions are \emph{not yet irreversibly ordered}, thus
opening up the possibility for front-running attacks.
Furthermore, under certain probabilistic consensus algorithms,
such as PoW or PoS, a transaction
inserted onto the blockchain can still be reordered by inducing
a fork of the \ub. 
Hence, to guarantee irreversible ordering for probabilistic consensus algorithms,
a transaction must receive enough block confirmations---the number of blocks
succeeding the block containing the transaction~\cite{nakamoto2008bitcoin,blockconfirmationsKraken,blockconfirmationsCoinbase}. 

\subsection{Smart Contract \& Decentralized Exchange}

A smart contract is an executable computer program modeled
after a contract or an agreement that executes 
automatically~\cite{savelyev2017contract}.
A natural fit for smart contracts is on top of 
decentralized fault-tolerant consensus algorithms, such as 
PBFT-style algorithms, PoW, and PoS, to
ensure their execution and integrity~\cite{wood2014ethereum,nakamoto2008bitcoin,kogias2016enhancing}.

Although Bitcoin uses a form of smart contracts~\cite{nakamoto2008bitcoin},
it was not until Ethereum's introduction that the blockchain space
realized Turing-complete smart contracts, the backbone necessary
for creating complex decentralized exchanges.
To interact with these complex smart contracts, users
need to pay \textit{gas}, a pseudo-currency that represents the execution
cost by miners~\cite{ethereumGas}.
However, the expressiveness of smart contracts comes with
significant risks, from inadvertent vulnerabilities to front-running.
Front-running is exhibited by the lack of guarantees
that the \ub provides regarding ordering.

\subsection{Front-Running Attacks \& Mitigation}\label{front-running-attacks}

The practice of front-running involves benefiting from advanced knowledge of pending
transactions~\cite{eskandari2019sok,bernhardt2008front,NasdaqFrontRunning}.
In itself, knowledge of pending transactions is harmless, but 
the ability to act on this information is where the true problem lies.
In the context of blockchains, an adversary performs a front-running attack 
by influencing the order of transactions, provided that
transactions in the mempool are entirely in the clear.

Cryptocurrencies suffer from mainly three types of front-running attacks~\cite{eskandari2019sok}: 
displacement, insertion, and suppression.
\emph{Displacement} is the replacement of a target transaction with a new transaction formulated by the front-running attacker.
\emph{Insertion} is the malicious introduction of a new transaction before a target transaction in the finalized transaction ordering.
\emph{Suppression} is the long-term or indefinite delaying of a target transaction.
%
%

In an ideal world, front-running protection would consist of an \emph{immediate}
global ordering of each transaction, as clients broadcast their transactions
to prevent attackers from changing their order.
In reality, even if all participants were honest, such global ordering is practically 
impossible due to clock synchronization~\cite{dalwadi2017comparative}
and consistency problems (\eg, two transactions having the
same time).
Malicious participants can still carry out front-running
attacks, because timings can easily be manipulated.

A more practical solution involves encrypting transactions,
thereby preventing the \cg 
from knowing the contents of the transactions 
when ordering them.
This solution mitigates front-running attacks as 
an attacker is hindered from taking advantage of pending \emph{encrypted} transactions.

\section{Strawman Protocols}

\label{sec:Strawman}

In order to explore the challenges inherent in building a 
framework, such as \system, 
we first examine a couple of promising 
but inadequate strawman approaches,
representative of state-of-the-art proposals~\cite{libsubmarine,breidenbach2018enter,momeni2022fairblock,shutter} 
but simplified for expository purposes.

\subsection{Strawman I: Sender Commit-and-Reveal}
\label{sec:StrawmanI}
The first strawman design has the sender create two transactions:
a \emph{commit} and a \emph{reveal} transaction.
The commit transaction is simply a commitment 
(\eg, hash) of the intended reveal transaction,
which is simply the typical contents of a transaction
that is normally vulnerable to front-running.
The sender will propagate the commit 
transaction and then \emph{wait} 
until its finality by the consensus group,
before releasing the reveal transaction.
Once the reveal transaction is propagated, 
the consensus group proceeds to verify and 
to execute the transaction, in the execution
order that the commit transaction 
was finalized on the blockchain.
Given the finality in the former transaction,
the sender is unable to change the contents
of the reveal transaction.

This simple strawman protocol mitigates front-running attacks
because the commit transaction determines the execution order
and the contents of the commit transaction
do not expose the contents of the reveal transaction.
However, this strawman protocol presents some notable challenges:
(a) the sender must remain online to 
continuously monitor the blockchain to know when
to release their reveal transaction,
(b) the reveal transaction might be delayed
due to a congestion event like the cryptokitties mania~\cite{CryptoKitties}
or a deliberate denial-of-service (DoS) attack like the Fomo3D incident~\cite{eskandari2019sok},
(c) this approach is subject to output bias,
as the consensus nodes or the sender 
can deliberately choose not to
reveal certain transactions
during the reveal phase~\cite{baum2021sok}, 
such as only revealing profitable ones and aborting others,
and
(d) this approach has a significant latency overhead 
of over 100\%, given that
the sender must now send two
non-overlapping transactions instead of the 
one standard transaction.
\subsection{Strawman II: The Trusted Custodian}
A straightforward method for removing the sender from the equation, after sending the commit transaction, is to employ a
trusted custodian.
After the consensus group finalizes the transaction onto the \ub,
the trusted custodian reveals
the transaction's contents.

This strawman protocol mitigates front-running attacks,
as the nodes cannot read, before ordering, the contents of the 
transaction.
However,
the trusted custodian presents a single point of failure:
Consensus nodes cannot decrypt and execute a transaction
if the custodian crashes.
Instead, by employing a \emph{decentralized} custodian,
we can mitigate 
the single point of failure issues.

\subsection{Strawman III: Threshold Encryption with Block Key}
\label{sec:StrawmanIII}

The next natural step is to have a decentralized committee
that generates a public key for each block, thus enabling a user
to encrypt their transaction for a future block.
The committee would then release
the private key after the block finality.
Furthermore, the committee can use 
identity-based encryption~\cite{shamir1984identity}
to enable users to derive a future block key based on the block's height.

This strawman protocol seems to mitigate front-running, as
the transactions in a block are encrypted until they are finalized in
their intended block.
However, if an encrypted transaction fails to be included in the specified 
block, 
its contents will be revealed shortly thereafter
while remaining unfinalized, thus making it vulnerable to front-running.
Blockchain networks have repeatedly observed such failures
due to congestion, 
such as cryptokitties manias~\cite{CryptoKitties}, 
or well-funded DoS attacks,
such as the Fomo3D attack
that flooded the Ethereum network with transactions
for three minutes~\cite{eskandari2019sok}.
Such an approach can incentivize
a \cn to intentionally produce an empty block
by aiming to reveal the pending transactions for that block.
Therefore, we require a per-transaction
 rather than a per-block level of confidentiality, thus
ensuring that a transaction is never revealed 
before it is finalized on the blockchain.
\section{System Overview}

In this section, we present \system's system goals,
architecture, and models.

\subsection{System Goals}
\label{sec:goals}
Our system goals, inspired by our strawman protocols, are
\begin{itemize}
    \item \textbf{Front-Running Protection:}
    prevents entities from practicing front-running.
    
    \item \textbf{Decentralization:} 
    mitigates a single point of failure or compromise.
    
    \item \textbf{Confidentiality:} 
    reveals a transaction, only after the underlying consensus layer 
	finalizes it.
    
    \item \textbf{Compatibility:}
    remains agnostic to the underlying consensus 
	algorithm and to smart contract implementation. 
    
    \item \textbf{Low-Latency}:
    exhibits low-latency transaction-processing overhead.
    
\end{itemize}

\subsection{Architecture Overview}
\System, shown in \Cref{Fig:architecture},
mitigates front-running attacks by working
with a \smc to manage the storage and release of
on-chain secrets.
Instead of propagating their transactions in cleartext,
the sender can now encrypt their transactions
and store the corresponding secret keys with the 
\smc.
Once the transaction is 
finalized, the \smc releases the secret keys
so that consensus nodes of the underlying blockchain
can verify and execute transactions.
Overall, the state machine replication
of the \ub is achieved in two steps: 
the first is about the ordering of transactions,
and the second is about the execution of transactions.
As long as most trustees in the \smc are secure and honest
and the key is revealed to the public when appropriate,
each consensus node can always maintain the same blockchain state.

\system encrypts 
the entire transaction\footnote{\Cref{sec:metadata} further discusses 
how to hide the sender's address.}, 
such as the smart contract address, inputs, 
sender's signature, and other metadata,
as those information
can provide enough information
to launch a probabilistic front-running attack,
such as the Fomo3D attack~\cite{eskandari2019sok} or 
a speculative attack based on the leakage of metadata~\cite{baum2021sok}.

\subsection{System and Network Model}

\system's architecture consists of three components:
\emph{senders} that publish (encrypted)
 transactions,
the \emph{\smc} (\SMCa) that manages and releases secrets,
and the \emph{consensus group} that maintains the 
\ub.
For the \system based on the PVSS scheme,
the client can choose a different \SMCa
for each transaction.
For the \system based on the THD2 scheme,
an \smca has a fixed membership over one epoch.
When transiting from one epoch to the next, 
the \smca can modify its membership under the THD2 scheme
with backward secrecy
to prevent new trustees from decrypting old transactions
without interrupting users’ encryption 
by running a resharing protocol~\cite{wong2002verifiable}.

The \smc and the \cg
can consist of the same set of servers.
For clarity in this paper, 
we logically separate them into two different entities.

For the underlying network,
we assume that all honest blockchain nodes 
and trustees of the \smca are well connected and
that their communication channels are synchronous, 
\ie, if an honest node or trustee broadcasts a message,
then all honest nodes and trustees receive the message 
within a known maximum delay~\cite{pass2017analysis}.

\subsection{Threat Model}
\label{sec:threatmodel}
We assume that the adversary is computationally bounded,
that the cryptographic primitives we use are secure,
and in particular that the Diffie-Hellman problem and its decisional variant are hard.
We further assume that all messages are digitally signed
and that the consensus nodes and the \smca only process
correctly signed messages.

The secret management committee consists of $n$ trustees, 
where $f$ can fail or behave maliciously.
We require
$n \geq 2f + 1$  and 
set the secret-recovery threshold to $t = f + 1$.
We assume that the underlying blockchain is secure:
\eg, at most $f'$ of $3f'+1$ 
validators can
fail or misbehave in a PBFT-style or PoS blockchain,
or the adversary controls less than $50\%$ computational power 
in a PoW blockchain.
We acknowledge that the security assumptions
for the secret management committee and the \ub 
might differ, 
potentially reducing the overall system's security to the least secure subsystem.

We assume that attackers do not launch 
speculative front-running attacks~\cite{baum2021sok}, 
but we present a discussion on some mitigation strategies 
for reducing side-channel leakage in \Cref{sec:metadata}.
\section{F3B Protocol}

\begin{figure}[t]
    \centering
    \includegraphics[scale=0.5]{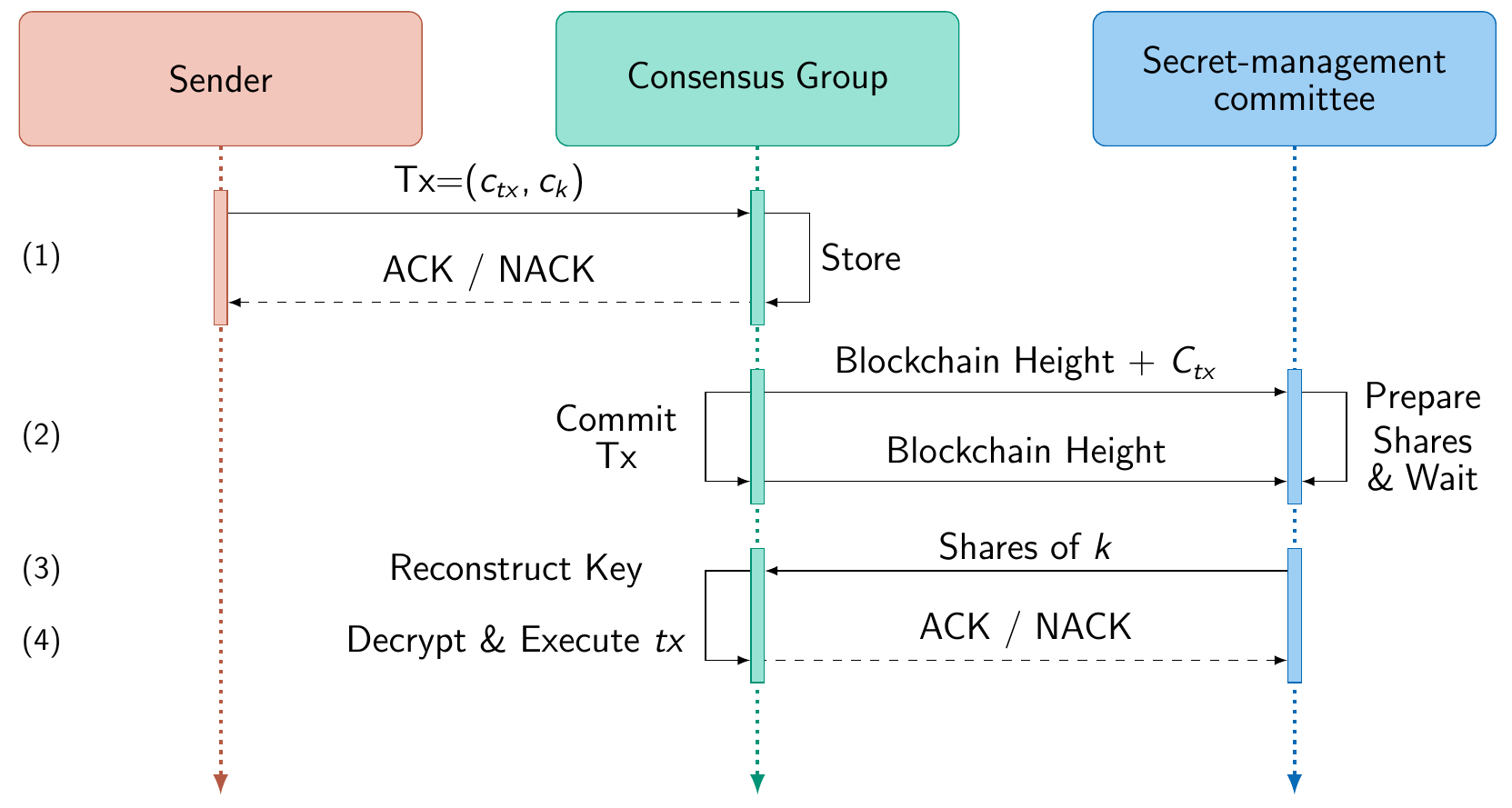}
    \caption{\System per-transaction protocol steps: 
    (1) Send an encrypted transaction to the underlying blockchain, 
    (2) Prepare shares by trustees while waiting for transaction finality,
    (3) Reconstruct the key,
    (4) Execute the transaction.
    }
    \label{Fig:protocol}
\end{figure}

In this section, we introduce the \system's protocol,
starting with some preliminaries, followed by the \system's detailed design.
\Cref{sec:fullprotocol}
offers a more comprehensive protocol description,
and \Cref{sec:discussion} introduces some optimizations.

\subsection{Preliminaries}

In this subsection, we introduce our preliminaries,
including our baseline model for the underlying blockchain
and the cryptographic primitives used in \system.

\paragraph*{Blockchain Model}
\label{sec:baseline}
To compare F3B's impact, we model the underlying
blockchain to involve a consensus protocol that finalizes
transactions into a block that is
linked to a previous block.
We assume the underlying block's time as $L_b$ seconds.
In PoW and PoS-based blockchains,
a transaction is finalized only
after a certain number of additional blocks 
have been added to the chain 
(also known as block confirmations).
Thus, we define that a transaction is finalized after $m$ block confirmations.
Therefore, our baseline transaction latency
is $mL_b$.

\paragraph*{Shamir's Secret Sharing}
A $(t, n)$-threshold secret sharing scheme
enables a dealer to share 
a secret $s$ among $n$ trustees such that
any group of $t \leq n$ or more trustees can reconstruct $s$ and
no group less than $t$ trustees learns any information about $s$.
Whereas a simple secret sharing scheme assumes an honest dealer,
verifiable secret sharing (VSS) enables the trustees to verify
that the shares they receive are valid~\cite{feldman1987practical}.
Public verifiable secret sharing (PVSS) further improves VSS
to enable a third party to check all shares~\cite{schoenmakers1999simple}.

\paragraph*{Distributed Key Generation (DKG)}
DKG is a multi-party $(t, n)$ key-generation process
for collectively generating a private-public key pair $(sk, pk)$,
without relying on a single trusted dealer;
each trustee $i$ obtains a share $sk_i$ of the secret key 
$sk$, and collectively obtains 
a public key $pk$~\cite{gennaro1999secure}.
Any client can now use $pk$ to encrypt a secret,
and at least $t$ trustees must cooperate to retrieve this
secret~\cite{shoup1998securing}.

\subsection{Protocol Outline}
\label{sec:protocol}

We present the outline of \System protocols with
two different threshold cryptographic schemes.
\Cref{Fig:protocol} presents the protocol outline,
and \Cref{sec:fullprotocol} offers a more comprehensive protocol description.

\subsubsection{Protocol based on TDH2}
\label{subsubsec:shortTDH2protocol}

\paragraph*{Setup:}
Before an epoch,
the \smc runs a DKG protocol to generate
a private key share $sk^i_{smc}$ 
for each trustee and a collective public key $pk_{smc}$
written onto the \ub.
To offer chosen-ciphertext attack protection 
and to verify the correctness of secret shares, 
we utilize the TDH2 cryptosystem~\cite{shoup1998securing} containing NIZK proofs.

\paragraph*{Per-Transaction Protocol:}

\begin{enumerate}
    \item \textit{Write Transaction:} 
    A sender first generates a symmetric key $k$
    and encrypts it with $pk_{smc}$ from the \ub, thus obtaining the 
    resulting ciphertext $c_k$.
    Next, the sender creates their signed transaction and symmetrically 
    encrypts it by using $k$, denoted as $c_{tx}=enc_k(tx)$.
    Finally, the sender sends $(c_{tx},c_k)$ to the consensus group
    who writes the pair
    onto the blockchain.
    \label{item:sharing}

    \item \textit{Shares Preparation by Trustees:}
    Once written, each \smc trustee reads $c_k$ from the sender's
    transaction and prepares their decrypted share of $k$.
    
    \label{item:sharestrustees}
    \item \textit{Key Reconstruction:}
    When the sender's transaction $(c_{tx},c_k)$ 
    is finalized onto the \ub
    (after $m$ block confirmations),
    each \smc trustee releases their share to the \cg.
    The \cg verifies the decrypted shares and 
    uses them to reconstruct $k$ by Lagrange interpolation 
    of shares when there are at least $t$ valid shares.
    \label{item:reconstruction}
    \item \textit{Decryption and Execution:}
    The \cg finally symmetrically decrypts the
    transaction $tx=dec_k(c_{tx})$ using $k$, thus 
    enabling it to execute $tx$.
    \label{item:execution}
\end{enumerate}

\paragraph*{Resharing Protocol:}
To modify a \SMCa's membership and to offer backward secrecy over epochs,
an \smca can periodically run a verifiable resharing protocol~\cite{wong2002verifiable}
to replace certain trustees or 
redistribute the trustees' private keys.
Unlike DKG, resharing keeps the epoch's public key, thus
preventing undesirable interruptions
of encryption services. 

\subsubsection{Protocol based on PVSS}
\label{subsubsec:shortFULLprotocol}

\paragraph*{Per-Transaction Protocol:}

\begin{enumerate}
    \addtocounter{enumi}{-1}
    \item \textit{Share Preparation By sender:}
    \label{item:sharepreparation}
    For every transaction, the sender runs the PVSS protocol \cite{schoenmakers1999simple} 
    to generate an encrypted key share $share_i$ for each trustee, 
    as well as a corresponding NIZK proof and public polynomial commitment. 
    The proof and commitment can be used to verify the correctness of key share 
    and protect against chosen-ciphertext attacks.
    The sender obtains the symmetric key $k$ from the PVSS protocol.

    \item \textit{Write Transaction:} 
    A sender first creates the ciphertext $c_k$ with the key shares, NIZK proofs, and commitments generated during share preparation.
    Next, the sender creates their transaction and symmetrically 
    encrypts it by using the symmetric key $k$, denoted as $c_{tx}=enc_k(tx)$.
    Finally, the sender sends $(c_{tx},c_k)$ to the consensus group
    who writes the pair
    onto the blockchain.

    \item \textit{Shares Preparation by Trustees:}
    Same as (\ref{item:sharestrustees}) in \ref{subsubsec:shortTDH2protocol}.

    \item \textit{Key Reconstruction:}
    Same as (\ref{item:reconstruction}) in \ref{subsubsec:shortTDH2protocol}.

    \item \textit{Decryption and Execution:}
    Same as (\ref{item:execution}) in \ref{subsubsec:shortTDH2protocol}.    
    
\end{enumerate}

\subsection{Overhead Analysis}
\label{sec:OverheadAnalysis}

We analyze both protocols' overheads.
Write Transaction (step 1) is identical 
to sending a transaction to the \ub.
We assume trustees can finish 
Shares Preparation by Trustees (step 2)
within the confirmation time of the \tx\footnote{
As we presented in \Cref{sec:evaluation},
confirmation time in Ethereum is much longer than the share preparation time by trustees.}.
Hence, the time for steps \ref{item:sharing} and \ref{item:sharestrustees}
is equivalent to finalizing a 
transaction on the underlying blockchain and waiting until 
its finality, which takes $mL_b$ time based on our 
baseline model (\Cref{sec:baseline}).
As in PVSS protocol,
the sender can finish Share Preparation By sender (step \ref{item:sharepreparation})
before having the $tx$;
thus step \ref{item:sharepreparation} does not contribute
to the transaction latency.
Comparing our protocol with the baseline,
Key Reconstruction (step \ref{item:reconstruction}) and Decryption and Execution (step \ref{item:execution})
are additional steps,
and we denote the time of those steps to be $L_r$.
\begin{figure}[t]
    \centering
    \begin{subfigure}[t]{0.495\textwidth}
    \includegraphics[scale=0.3]{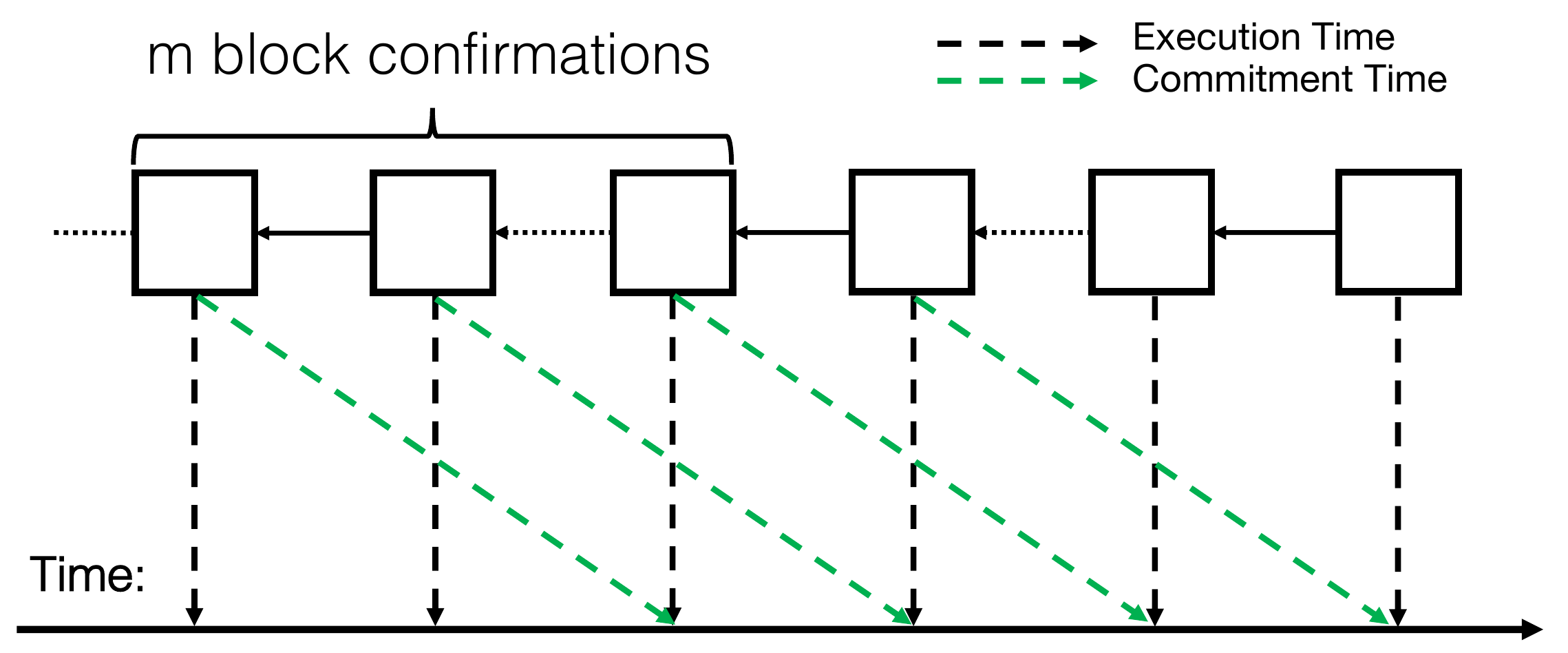}
    \caption{Execution and finality time in Ethereum} \label{fig:execitiona}
    \end{subfigure}
    \begin{subfigure}[t]{0.496\textwidth}
    \includegraphics[scale=0.3]{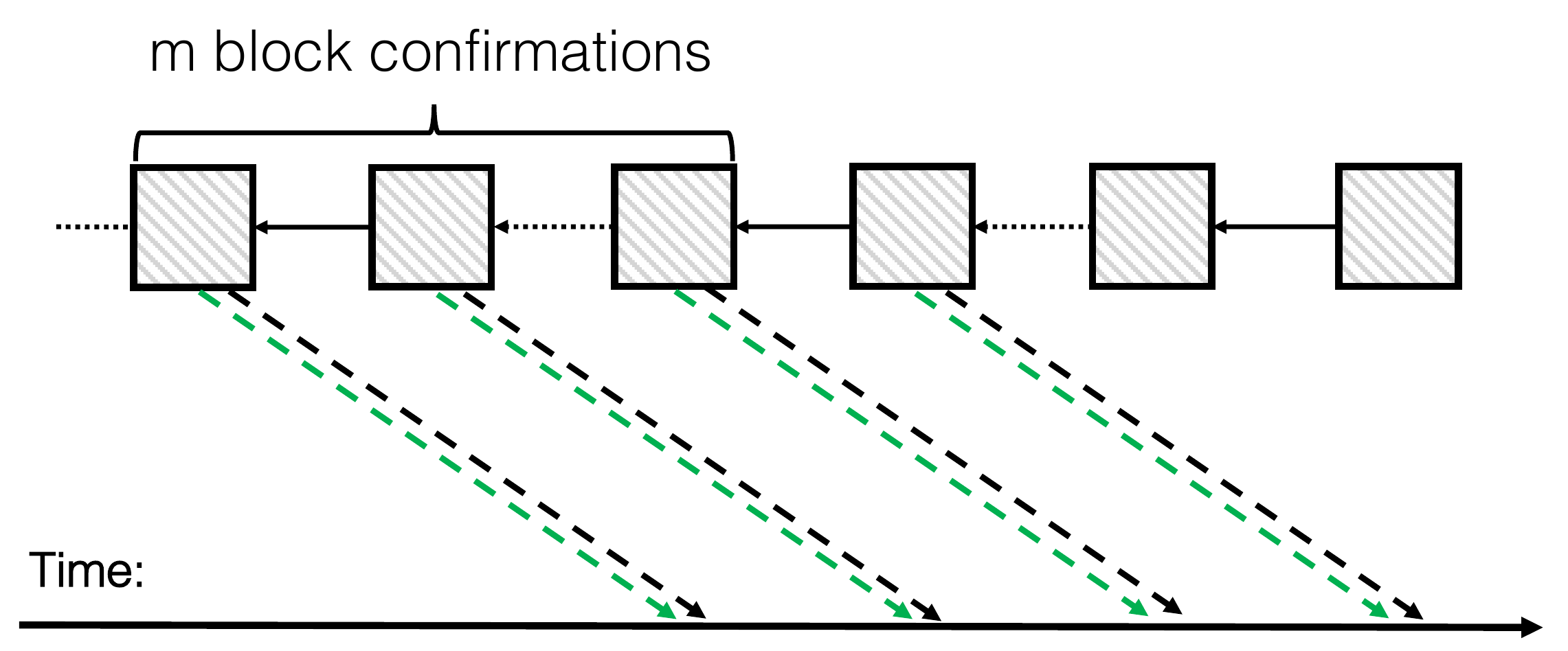}
    \caption{Execution and finality time in F3B} \label{fig:execitionb}
    \end{subfigure}
    \caption{In Ethereum, 
    once they are inserted in the blockchain, the transactions are executed and finalized after receiving $m$ block confirmations.
    Whereas, in \system, transactions are encrypted, and their executions are postponed
    after receiving $m$ block confirmations when
    the \smc releases the encryption keys.
    Both scenarios have a similar finality time.
    } \label{fig:execition}
\end{figure}

\Cref{fig:execition} demonstrates
the conceptual difference in finality and execution time between 
F3B and the baseline.
As the \smc releases the secret keys with a delay of $m$ blocks,
this introduces an execution delay of $m$ blocks.
However, in both cases, to prevent attacks
such as double-spending, the recipient should not accept a 
transaction, until it is finalized.
Therefore on a commercial level, \system is similar to
the baseline 
because it exhibits the finality time of a transaction
that is of use to the recipient\footnote{In \system, 
transaction finalization is slower due to the key reconstruction
and delayed execution after transaction finality.
However, the overhead is negligible compared to finality time,
as discussed in \Cref{sec:evaluation}.}.

\subsection{TDH2 and PVSS: Pros and Cons}
\label{sec:comparsion}
When applying THD2 and PVSS to \system, 
each scheme has some advantages and disadvantages.
This subsection offers qualitative comparisons,
whereas \Cref{sec:evaluation} provides quantitative comparisons 
between the two protocols. 

\begin{itemize}
    \item \textbf{Preprocessing:}
    In TDH2, the \smc needs to do DKG per epoch,
    whereas in PVSS, 
    the sender needs to prepare shares per transaction.
    
    \item \textbf{Membership:} 
    In TDH2, the \smc's membership is fixed per epoch,
    whereas in PVSS, 
    the sender can choose a different \smc for each transaction,
    providing the best flexibility.

    \item \textbf{Ciphertext:} 
    TDH2 has a constant ciphertext length,
    whereas PVSS's ciphertext grows linearly with the size of the \smc.
    
\end{itemize}

In conclusion,
no one protocol can completely replace another.
System designers need to 
choose one or both protocols based on their needs and constraints
to mitigate front-running attacks.

\section{Achieving the System Goals}
\label{sec:achieving_goals}
In this section, we present how \system achieves the system goals outlined in 
\Cref{sec:goals}.

\textbf{Front-Running Protection:}
\emph{prevents entities from practicing front-running.}
    
We reason the protection offered by \system
from the definition of front-running:
if an adversary cannot benefit from pending transactions,
he cannot launch front-running attacks.
In \system, the sole entity that knows the content of a 
pending transaction is the sender who is
financially incentivized to \emph{not} release its contents.
The content is revealed only when its transaction is finalized;
thus, by definition, the attacker has no means to
launch a front-running attack.
However, we acknowledge that attackers 
can use side channels (\eg metadata such as
sender's address and transaction size) of
the encrypted transaction to launch \emph{speculative}
front-running attacks, as discussed 
in~\Cref{sec:threatmodel} and~\Cref{sec:metadata}.
We present a more comprehensive security analysis discussion in \Cref{sec:securityAnalysis}.
    
\textbf{Decentralization:} 
\emph{mitigates a single point of failure or compromise.}
    
Due to the properties of DKG~\cite{gennaro1999secure}, THD2~\cite{shoup1998securing}, and
PVSS~\cite{schoenmakers1999simple},
the \smca can handle up to $t-1$ malicious trustees
and up to $n-t$ offline trustees. 

\textbf{Confidentiality:} 
\emph{reveals a transaction, only after the underlying consensus layer finalizes it.}

The sender encrypts each transaction with a newly generated
symmetric key. The symmetric key is 
(a) encrypted under the \smc's public key in TDH2-based protocol, 
(b) embedded into the encrypted shares in PVSS-based protocol. 
In both protocols, $f+1$ trustees are required to retrieve the symmetric key.
Per our threat model, only $f$ trustees can behave maliciously; this ensures that the symmetric key cannot be revealed.
We outline a more detailed security analysis in \Cref{sec:securityAnalysis}.

\textbf{Compatibility:}
\emph{remains agnostic to the underlying consensus 
	algorithm and to smart contract implementation.}

\system requires modifying the execution layer to enable
encrypted transactions. However, the consensus layer
remains untouched, thus agnostic to the underlying 
consensus algorithms.
Furthermore, \system does not require to modify smart contract implementations,
thus enabling existing smart contracts to 
benefit from front-running protection automatically.
    
\textbf{Low-Latency:}
\emph{exhibits low-latency transaction-processing overhead.}

Similar to the baseline model,
F3B requires clients to write only one transaction 
onto the \ub.
This enables F3B to have a low-latency overhead
compared to other front-running protection design that 
require multiple transactions for the same security guarantees.
We present an evaluation of this latency overhead in~\Cref{sec:evaluation}.

\section{Security Analysis}
\label{sec:securityAnalysis}

In this section, we introduce the security analysis of \system's protocol.

\subsection{Front-Running Protection}

From our threat model, we reason about why an attacker can no longer
launch front-running attacks with absolute certainty of a financial
reward,
even with the collaboration of at most $f$ malicious trustees.
As we assume that the attacker does not launch speculative attacks
based on metadata of the encrypted transactions,
the only way the attacker can front-run transactions
is by using the plaintext content of the transaction.
As the attacker cannot access the content of the transaction
before it is finalized on the \ub, then
the attacker cannot benefit from the pending transaction.
This prevents front-running attacks (by the definition of front-running).
As we assume that the symmetric encryption we use is secure,
the attacker cannot decrypt the transaction based on its ciphertext.
Due to  the properties of 
TDH2~\cite{shoup1998securing},
DKG~\cite{gennaro1999secure}, and
PVSS~\cite{schoenmakers1999simple} with our threat model,
the attacker cannot obtain the private key and/or
reconstruct the symmetric key.
Recall that the attacker can  collude with at most only $f$ trustees, 
and that $f+1$ are required to recover 
or gain information about the symmetric key.

\subsection{Replay Attack}
\label{sec:ReplayAttack}
We consider a scenario in which
an adversary can copy a pending (encrypted) transaction 
and submit it as their own transaction to reveal the transaction's contents,
before the victim's transaction is finalized.
By revealing the contents of the copied transaction, 
the attacker can then trivially launch a front-running attack.
However, we explain the reason the adversary is unable to benefit from 
such a strategy.

In the first scenario, 
the adversary copies the ciphertext $c_k$
and the encrypted transaction $c_{\tx}$ from \txw,
then creates a new write transaction \txwp, 
digitally signed with their signature. 
However, even if the adversary's \txwp 
is decrypted and executed 
before the victim's transaction \txw, 
it effectively results in the blockchain executing \txw
\footnote{Note that \tx is already a signed transaction; 
thus, \txw and \txwp have the same effect on the blockchain.}.
This leaves the adversary with no time to front-run their own \txwp without knowing its contents.

In our second scenario, the adversary instead sends the
transaction to a blockchain with smaller $m$ block confirmations.
Consider two blockchains $b_1$ and $b_2$
whose required number of confirmation blocks are $m_1$ and $m_2$
with $m_1 > m_2$.
If the adversary changes the label $L$ to $L'$ for the blockchain 
$b_2$ instead of blockchain $b_1$,
the \smc will successfully decrypt the transaction.
However, we argue it is hard to form a valid write-transaction 
with $L'$ by the adversary.

For the TDH2 protocol,
the adversary would need to generate
$e'=\hash_1\left(c,u,\bar{u},w,\bar{w},L'\right)$
and $f=s+re'$,
without knowing the random parameter $r$ and $s$.
Suppose the adversary generates 
$u=g^r, \bar{u}=\bar{g}^{r'}$ with $r \neq r'$ and 
$w=g^s, \bar{w}=\bar{g}^{s'}$ with $s \neq s'$.
For \txwp to be valid, 
we must have $g^f = wu^e$ and $\bar{g}^f = \bar{w}\bar{u}^e$,
this implies that $(s-s')+e(r-r')=0$.
As $r \neq r'$ , the adversary has only a negligible chance of having
\txwp pass verification.

For the PVSS protocol, 
the adversary must replace the original generator $h$ with $h'$ 
derived from $H(L')$. 
Hence, the adversary has to do the proofs without secrets.
The security of PVSS guarantees that
they only have a negligible probability of succeeding.
Note that the base point has to be random
to ensure the security.
Using Elligator maps~\cite{bernstein2013elligator} guarantees that
the generator $h$ is random.
\section{Incentive}
\system must incentivize actors to operate
and follow the protocol honestly.
In this section, we address the critical incentives that, in \system,
prevent spamming transactions and that deter
collaboration among trustees from prematurely revealing transactions.

\subsection{Spamming Protection}
\label{sec:spamming}
As the \cg cannot execute encrypted transactions,
an adversary could, at a low cost, 
spam the blockchain with non-executable transactions 
(\eg, inadequate fees, malformed transactions), thus
delaying the finality of honest transactions.
To make such an attack costly, we introduce a storage deposit,
alongside the traditional execution fee (\eg, gas in Ethereum) and adjustable based on the transaction's size.
The \ub can deduct the storage deposit from the sender's balance, 
much like paying a transaction fee.
Then the \bc can partially refund the deposit 
after successful execution by the \cns. 
This approach imposes a low-cost fee on compliant users 
and a penalty on those who misbehave.

\subsection{Operational Incentive}\label{subsec:lengthofepoch}
We need a similar incentive structure for the \smcs; similar to the way consensus nodes are rewarded
for following the blockchain protocol via an execution fee.
Whereas {\smca}s could be rewarded using the execution fee, 
this fee does not prevent \smca trustees from colluding
for their financial gain.
For example, an \smca might silently collude with
a \cg by prematurely giving them the decryption shares.
Given the difficulty of detecting out-of-band collusion, 
we need to discourage it from doing so by significantly rewarding
anyone who can prove the existence of such collusion.

We propose an incentive structure, where
we require each trustee in a \smc to lock an amount $c$ for collateral
and, in exchange, they are rewarded proportionally to the staked amount 
$ac$ for the services they provide.
Remind that, based on our threat model (\Cref{sec:threatmodel}),
$t$ trustees must collude altogether to reconstruct a transaction.
To maintain security, 
the potential gain that $t$ trustees benefit 
from front-running must be less than
the potential loss $(1+a)ct$, 
which malicious trustees would incur 
through the slashing protocol described in 
\Cref{sec:protocol-incentive}. 
Hence,
a higher potential loss value $(1+a)ct$ ensures
security for a longer epoch length. 
If other factors stay consistent, 
designers can support a longer epoch length 
by either increasing the collateral requirement 
(by raising $c$) or 
by involving more trustees 
(by increasing $t$).

\begin{figure}[t]
    \centering
    \begin{minipage}[t]{.48\textwidth}
  \centering
    	\includegraphics[scale=0.42]{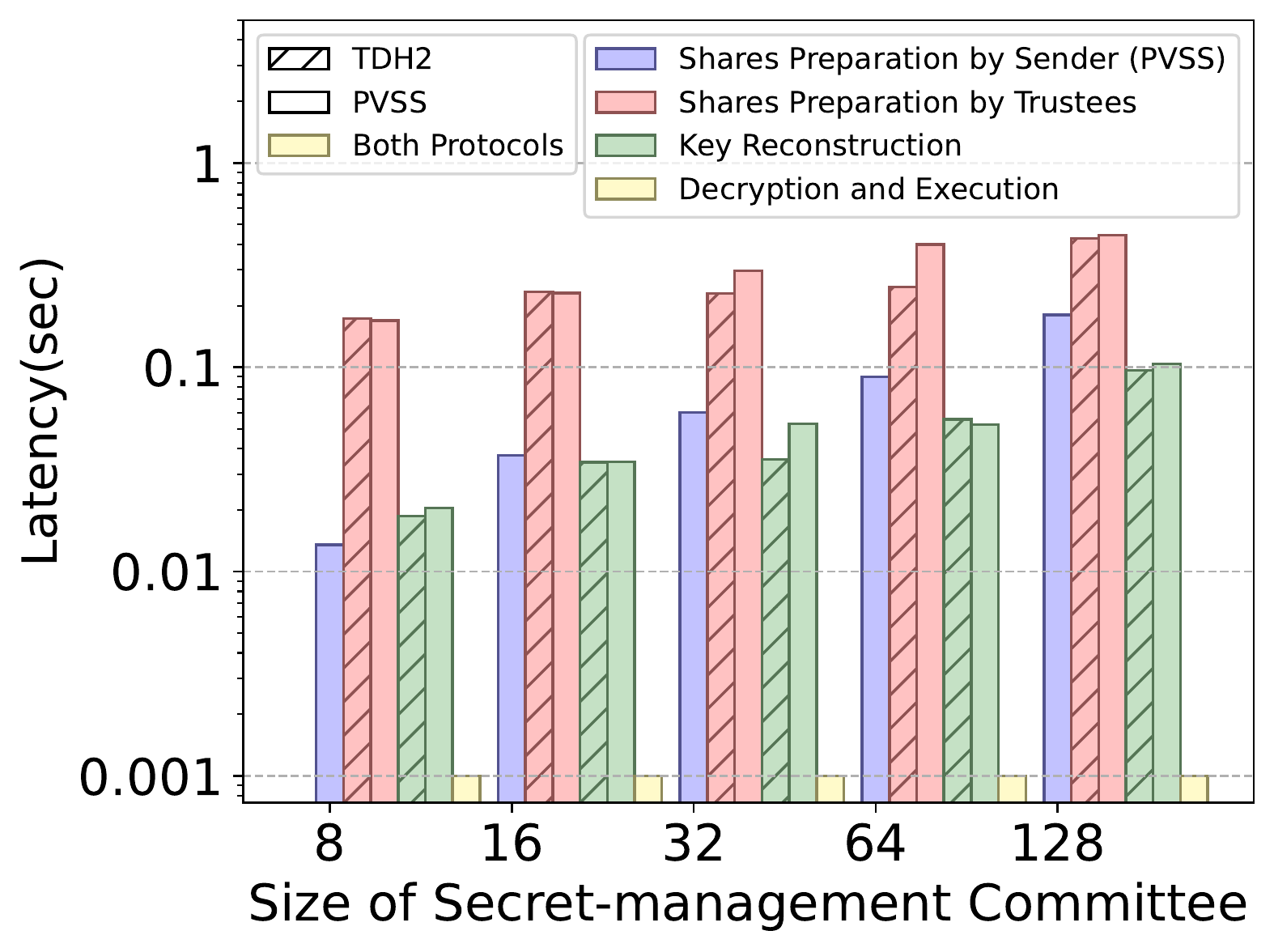}
    	\caption{The breakdown latency of each step in \system
    	by varying the number of \smc trustees from 8 to 128 nodes.}
    	\label{Fig:latency}
\end{minipage}\hfill
\begin{minipage}[t]{.48\textwidth}
  \centering
    \includegraphics[scale=0.42]{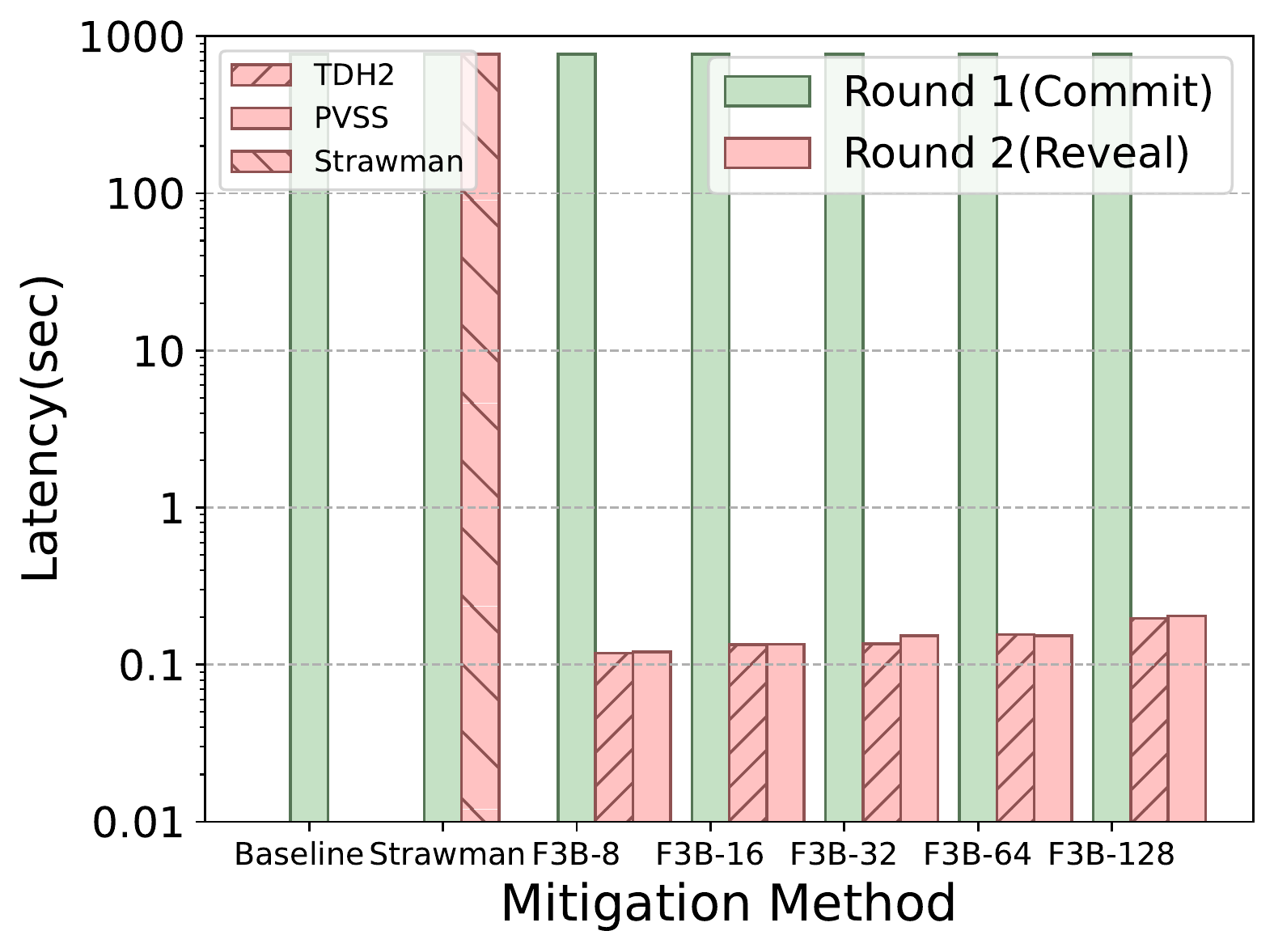}
    \caption{A comparison
        of the sender commit-and-reveal approach latency
        with \System
        against a baseline modeled in Ethereum.
        The string ``F3B-$X$'' represents $X$ trustees.  
    }
    \label{Fig:comparison}
\end{minipage}
 \end{figure}

\subsection{Slashing Protocol}
\label{sec:protocol-incentive}

We need to have a protocol that
rewards anyone who can prove a trustee's or 
the entire \smc's misbehavior 
to discourage the release of the shares prematurely.
At the same time, we do not want anyone to accuse
a \smc or a specific trustee without repercussions
if the \SMCa or trustee did not actually misbehave.

To accomplish our objective, each trustee of the \smc 
must stake some amount of cryptocurrency in a smart
contract that handles disputes between a defendant
(the entire \smc or a particular trustee) 
and a plaintiff.
To start a dispute, the plaintiff will invoke the smart
contract with the correct decryption share for a 
currently pending transaction and their own stake.
Suppose the smart contract validates that this is a correct
decryption share and that the dispute started before
the transaction in question was revealed by the \smc.
In this case, the defendant's stake is forfeited and 
sent to the plaintiff.

At a protocol level, to prove a correct decryption share in protocol with TDH2,
the plaintiff submits $[u_i, e_i, f_i]$ such that
$e_i=\hash_2\left(u_i,\hat{u_i},\hat{h_i}\right)$
where
$\hat{u_i}=\frac{u^{f_i}}{{u_i}^{e_i}}$ and 
$\hat{h_i}=\frac{g^{f_i}}{{h_i}^{e_i}}$.
In the protocol based on PVSS, 
the plaintiff submits  $[s_i, \pi_{s_i}]$, 
where $\pi_{s_i}$ is the NIZK proof that shows $\log_g pk_i = \log_{s_i}\hat{s_i}$.
Even if the sender knows $s_i$, 
it is impossible to maliciously slash a trustee without the $\pi_{s_i}$,
which only the corresponding trustee knows.

Deploying such a mechanism would
require the smart contract to access the ciphertext 
of a transaction (\eg, $u$ or $\hat{s_i}$ is necessary to verify the submitted share).

\section{Evaluation}
\label{sec:evaluation}

\begin{table}[t]
\centering 
\begin{minipage}[t]{.49\textwidth}
\centering
\resizebox{\textwidth}{!}{%
\begin{tabular}[t]{Scccc}
\toprule
& \multicolumn{4}{c}{Latency Overhead varying SMC sizes} \\ \cmidrule{2-5}
 & \multicolumn{2}{c}{TDH2} & \multicolumn{2}{c}{PVSS}  \\ 
Confirmations & 64 & 128 & 64 & 128 \\
                   \midrule
8 &  0.164\%  & 0.206\% & 0.160\%  & 0.214\% \\ 
16 & 0.082\%  & 0.103\% & 0.080\%  & 0.107\% \\ 
32 & 0.041\%  & 0.052\% & 0.040\%  & 0.053\% \\ 
\textbf{64\hspace{12pt}} & \textbf{0.020\%}  & \textbf{0.026\%} & \textbf{0.020\%}  & \textbf{0.027\%} \\ 
128 & 0.010\% & 0.013\% & 0.010\%  & 0.013\% \\
\bottomrule
\end{tabular} }
\caption{Latency Overhead for Ethereum Blockchain}
\label{tab:latency} 
\end{minipage}
\begin{minipage}[t]{.49\textwidth}
\centering
\resizebox{\textwidth}{!}{%
\begin{tabular}[t]{SSS}
\toprule
& \multicolumn{2}{l}{Storage Overhead (bytes)} \\ \cmidrule(l){2-3} 
{Number of trustees} & {TDH2 Protocol}              & {PVSS protocol}              \\ 
\midrule
8                           & 80                        & 792                       \\
16                          & 80                        & 1568                       \\
32                          & 80                        & 3120                      \\
64                          & 80                        & 6224                      \\
128                         & 80                        & 12432                      \\ \bottomrule
\end{tabular} }
\caption{Storage overhead for two protocols with different Secret-management Committee sizes.}
\label{tab:storage}
\end{minipage}

\end{table}

We prototype \system 
by using post-merge Ethereum~\cite{ethereumMerge} as the \ub
and Dela~\cite{Dela} written in Go~\cite{Go}
as the \smc for our evaluation.
Remaining consistent with Ethereum's  security assumptions,
one epoch length lasts 6.4 minutes and the trustees forming
the \smc from validators for a given epoch are randomly selected. 
We instantiate our cryptographic primitives by using the
Edward25519 elliptic curve with 128-bit security
supported by Kyber~\cite{kyber}, 
an advanced cryptographic library.
We ran experiments on a server
with 32GB of memory and 40 (2.1GHz) CPU cores. The network communication delay is simulated to be a fixed 100ms.
We further discuss \system integration with Ethereum in \Cref{sec:discussion}.

\subsection{Latency}

In \Cref{Fig:latency},
we present the breakdown latency of each step
for both TDH2 and PVSS protocols
after a transaction finality
while varying the number
of \smca trustees from 8 to 128 nodes:
(a) shares preparation by trustees, and
(b) key reconstruction, and
(c) decryption and execution.
In addition, 
we show the time needed for PVSS shares generation 
by the sender in purple of \Cref{Fig:latency}.
As discussed in \Cref{sec:OverheadAnalysis},
only (b) and (c) represent the overhead at the per-transaction level.

Recall that the overall transaction latency using
\system is $mL_b+L_r$ (\Cref{sec:protocol}).
In post-Merge Ethereum,
the block time is fixed to 12 seconds, \ie, $L_b=12$~\cite{ethereumBlock},
and, by official standard,
a block requires 64 block confirmations (two epochs)
to be ``finalized", \ie, $m=64$~\cite{ethereumPoS}.

\Cref{Fig:comparison} presents the end-to-end latency comparison
between the baseline protocol (\Cref{sec:baseline}), 
a sender-only commit-and-reveal protocol,
as presented in Strawman 1 (\Cref{sec:StrawmanI}), 
and \system's
 protocol---varying the size of the \smc stated after the string ``F3B-''.
With the new PoS consensus,
finalizing any data in Ethereum 
requires $mL_b=64*12=768$ seconds.
The baseline protocol's total latency is $768$ seconds, as it
requires only one write to the blockchain.
Recall that in the sender-based commit-and-reveal approach (Strawman I), the sender commits a hash to the blockchain, taking $768$ seconds, then reveals the transaction in another $768$ seconds, totaling $1536$ seconds. This results in a 100\% latency overhead compared to the baseline, as the two steps must be sequential: the hash must be finalized on the blockchain before the reveal transaction can be propagated. 
Submarine, a more advanced approach that conceals the smart contract address, requires three sequential transactions. The sender must publish these three transactions in order, with the blockchain finalizing each one before the next one can be sent, suffering a latency delay of $768*3 = 2304$ seconds or a 200\% latency overhead compared to the baseline~\cite{libsubmarine,breidenbach2018enter}.

Compared with \system,
the reveal phase (key-reconstruction step) 
does not require the sender to write any data onto the blockchain.
Therefore, we emphasize a significant difference between \system and 
other application-based commit-and-reveal approaches,
where \system  requires sending only one transaction to the \ub.
\Cref{Fig:comparison} shows that our design brings a low-latency overhead
of 197ms and 205ms for two protocols, equivalent to 0.026\% and 0.027\% for Ethereum
(relative to the 768 seconds finality time),
under an \smca size of 128.

We acknowledge that some Ethereum users may accept a lower 
confirmation number to accept a transaction,
even though Ethereum officially requires 64 blocks~\cite{ethereumPoS}.
Without loss of generality, 
we outline different confirmation numbers with \system's latency overhead
in \Cref{tab:latency}.

\begin{figure}[t]
\centering
\begin{minipage}[t]{.49\textwidth}
  \centering
  \includegraphics[width=1\linewidth]{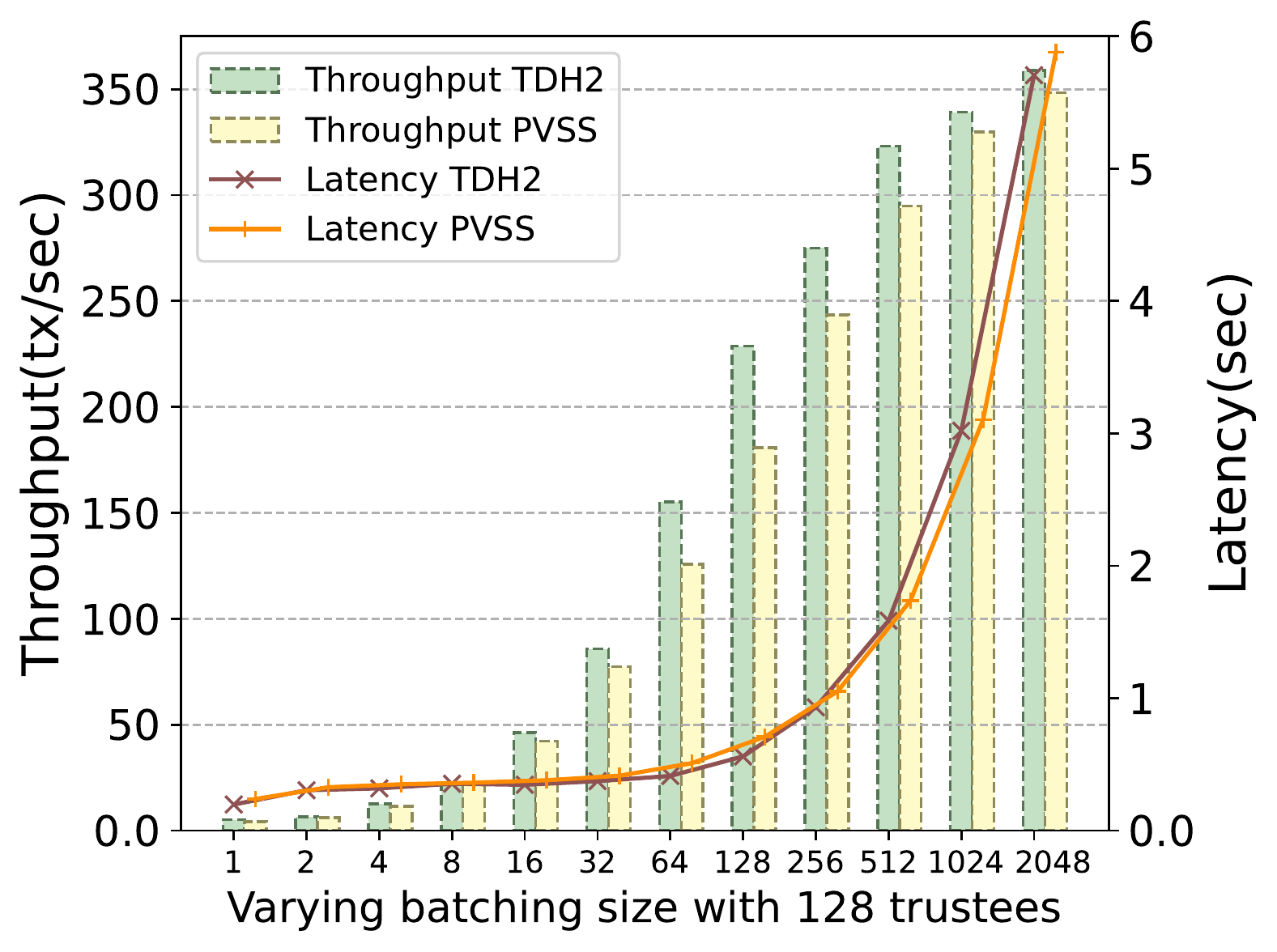}
    \caption{Performance of the key construction}
    \label{Fig:throughput}
\end{minipage}\hfill
\begin{minipage}[t]{.49\textwidth}
  \centering
     \includegraphics[width=1\linewidth]{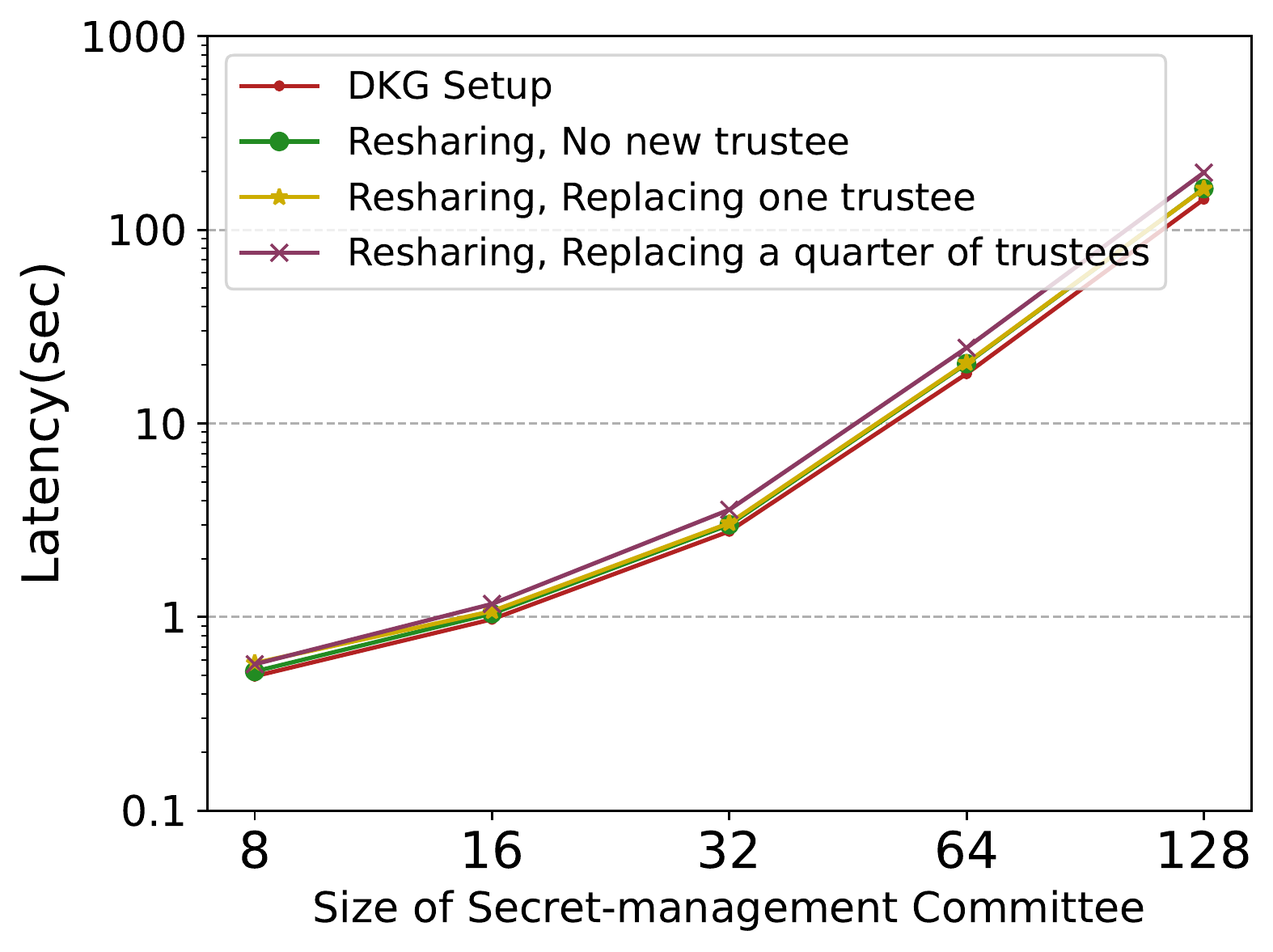}
    \caption{The latency cost of DKG setup and
    three resharing scenarios.}
    \label{Fig:resharing-latency}
\end{minipage}
\end{figure}
\subsection{Throughput}
\label{subsec:throughput}

\Cref{Fig:throughput} presents the \System's throughput results
with a \smc consisting of 128 trustees,
assuming the \ub is not the bottleneck.
If the keys are individually reconstructed,
\system provides limited throughput due to network
transmission overhead incurred from
sequential execution.
Instead, we can batch keys by reconstructing them
concurrently and presenting them in one network transmission.
We present this batching effect in \Cref{Fig:throughput} by
varying the batching size
to measure throughput and corresponding latency.
By increasing the batching size from 1 to 2048,
we can improve throughput from 5 txns/sec to 359 txns/sec with the TDH2 protocol,
and from 4 txns/sec to 348 txns/sec with the PVSS protocol.
The increased throughput comes with a
higher latency cost: With a batching size of 2048, 
the key reconstruction step of TDH2 now takes 5.71 seconds to process, 
and the same step of PVSS takes 5.88 seconds;
this latency is equivalent to a  0.74\% and 0.77\% latency overhead over Ethereum.
Our results show that \system provides more than sufficient throughput
to support Ethereum (15 tx/sec~\cite{schaffer2019performance}).

\subsection{Reconfiguration in TDH2}
\label{sec:resharing}

\Cref{Fig:resharing-latency} demonstrates the cost 
of reconfiguring a \smc in the TDH2 protocol.
Recall that DKG is a one-time setup operation per epoch,
bootstrapped during the previous epoch.
Our experiment shows that, 
with a committee size of 128 trustees,
DKG takes about 144 seconds;
this is about 37.5\% of Ethereum's epoch time (384 seconds).
We offer a further discussion 
about the transition between two epochs in \Cref{sec:transition}.
To provide backward secrecy and dynamic membership,
a \smc can run a 
verifiable resharing protocol~\cite{wong2002verifiable}
within an epoch and, by keeping the public key, without interrupting users' encryption.
\Cref{Fig:resharing-latency} illustrates the cost of three scenarios:
(a) resharing among the same committee,
(b) replacing one trustee, and
(c) replacing a quarter of trustees.
They all exhibit latency of the same magnitude.

\subsection{Storage Overhead}
\label{sec:storage}

In the TDH2 protocol, 
as the symmetric key is encrypted with the shared public key, 
the size of $c_k$ is independent of the number of trustees. 
We also optimize the original TDH2 protocol to 
remove the label $L$ from the ciphertext
but only insert $L$ in the computation and verification steps
of each party (\cg, \smc, sender) 
for protection against replay attacks (\Cref{sec:ReplayAttack}).
Ultimately, we achieve 80 bytes per transaction 
of the storage overhead presented in \Cref{tab:storage}.

In the PVSS protocol, however,
the ciphertext $c_k$ contains encrypted shares, NIZK proofs, and polynomial commitments. 
The size of the $c_k$ thus approximately grows linearly with the number of trustees,
as demonstrated in \Cref{tab:storage}.
The difference in storage overhead is one of the trade-offs 
when system designers need to consider using 
which encryption algorithm in \system, 
as discussed in \Cref{sec:comparsion}.

\section{Discussion}
\label{sec:discussion}

In this section, we discuss some deployment challenges.
We leave a detailed analysis for future work.

\subsection{Transition of Epoch}
\label{sec:transition}
Each epoch has its unique public epoch key for a user to encrypt
his symmetric key used for encrypting a transaction.
However, users will have difficulty choosing 
the correct epoch key
when the time is close to the transition between two epochs.
With Fairblock~\cite{momeni2022fairblock} and Shutter~\cite{shutter,shutterL1},
undesirable transaction revealing occurs
when a user chooses the wrong public key. 
Whereas, if the transaction is not finalized, \system never reveals any transaction, regardless of the chosen key; thus offering confidentiality to all unfinalized transactions.
We also expect that such an epoch transition is infrequent,
compared with a block transition,
thus it causes much less trouble to users.
In \system, if a user uses an old epoch key for his encryption,
he can safely try again to select the new epoch key.
To mitigate the issue even more, 
the expiring epoch committee can offer some grace period, thus 
allowing both old and new epoch keys to be valid for a certain period. 
This significantly reduces the danger of a user choosing an incorrect key.

\subsection{Ethereum Gas Fees}
\label{sec:EthereumGas}
Ethereum uses gas fees to cover the cost of executing a transaction
and implements a maximum gas limit per block.
Incorporating \system on Ethereum would then require 
(1) that the gas limit of each transaction to be in cleartext, and (2) that the summation of all transactions' \textit{gas limit} within
a block does not exceed the \textit{block gas limit}.
This opens the possibility for another type of
spamming attack (\Cref{sec:spamming}),
where an adversary submits transactions with
substantial gas limits, thus leaving little room for other
transactions.

Recall that the actual gas used by transactions cannot be 
determined because the sender encrypts its contents, and that accurately estimating the gas cost of a transaction is particularly difficult due to the uncertainty of the global state when validators process the transaction.
One potential approach to mitigating this kind of attack
would be to then burn the remaining (unused) gas.
However, this approach could be too strict in practice, hence
we can instead envision partial refunds: refunding the remaining
gas up to a percentage.

\subsection{Verifiable Key Propagation}
Under our proposed protocol, 
every consensus node must fetch the 
shares and run the Lagrange interpolation 
to reconstruct the key.
Would it instead be possible for one of the consensus nodes to reconstruct
the symmetric key $k$ from the secret shares and to propagate it
to other consensus nodes with a succinct proof?

Therefore, we propose a solution that requires additional storage overhead in exchange for faster verification:
Instead of constructing their encrypted transaction as 
($c_k, c_{\tx}$), the sender additionally adds a hash
of the symmetric key $h_k=H(k)$ as the third entry,
creating the following signed write transaction:
$\txw = [c_k, c_{\tx}, h_k]_{\sign{\sk_A}}$.

During key reconstruction,
after recovering or receiving $k$,
consensus nodes need to verify
whether the hash of $k$ is consistent with the one ($h_k$)
published on the ledger.
If it is consistent,
a \cn can continue to decrypt the transaction
and propagate the key $k$ to others.
If it is inconsistent, however,
a \cn must reconstruct the key from decryption shares
and publish the shares to the \ub to slash the sender
who provides a wrong $h_k$.

\subsection{Metadata Leakage}
\label{sec:metadata}
In our architecture, adversaries can only observe encrypted 
transactions until they are finalized, 
thus preventing the revelation of transaction contents
to launch front-running attacks.
Nevertheless, to launch speculative
attacks, adversaries can rely on side channels
such as transaction metadata.
Concretely, as the sender needs to pay the storage 
fee (\Cref{sec:spamming})
for publishing an encrypted transaction to the underlying
blockchain, this leaks the sender's address.
Knowledge of the sender's address can help in 
launching a front-running attack because an adversary might be able,  based on the sender's history, to 
predict the sender's behavior.
To prevent this second-order front-running attack,
a sender can use a different address to pay for the storage fee.
The \ub can also offer anonymous payment
to users, such as Zerocash~\cite{sasson2014zerocash}
or a mixing service~\cite{ziegeldorf2018secure},
to further hide the payment address.
Another side-channel leakage is the size of the encrypted
transaction or the time the transaction is propagated.
A possible remedy for mitigating metadata leakage
is PURBs~\cite{nikitin2019reducing}.

\subsection{Key Storage and Node Catchup}
In our protocol,
if a new node wants to join the consensus group,
it cannot execute the historical transactions to catch up,
unless it obtains all decryption keys.
The \smc or consensus group can store these keys
independently from the blockchain,
but this requires them to maintain an additional immutable ledger.
As consensus nodes already maintain one immutable storage medium, namely the underlying blockchain,
the keys can be stored on this medium as metadata;
and the blockchain rule can require storing valid keys when producing
blocks.

However, this optimization brings about a timing issue, \ie,
When should the blockchain require the consensus group to store
keys in a block?
From our protocol, the transaction is finalized at block height $n$ and
revealed at block height $n+m$, thus
making the earliest block to write the key at block height $n+m+1$.
With respect to the latest block height to write the key, there is much
more flexibility and we need to consider the balance between
the delay tolerance for all consensus nodes to retrieve the key and
the time that consensus nodes must retain the key.
Assuming that the key reconstruction step takes up to $\delta$ block times,
the key should be written in or before the block $n+m+\delta$.

Although this setup would work well for a blockchain with fixed block time,
care must be taken for blockchains where block time is probabilistic
as the key might not have been replicated to all consensus nodes 
at block height $n+m+\delta$, thus  some artificial delay for new blocks could be induced.

\section{Related Work}

Namecoin is a decentralized name service and
an early work on front-running protection using a commit-and-reveal 
design~\cite{kalodner2015empirical}.
In Namecoin, a user first broadcasts a salted hash of their name
and then, after finality, broadcasts the actual name.
Our first strawman protocol (\Cref{sec:StrawmanI}) is based on Namecoin.

After Namecoin, Eskandari et al.~\cite{eskandari2019sok} systematized 
front-running attacks on the \bc by
presenting three types of front-running attacks:
displacement, insertion, and suppression.
Daian et al.~\cite{daian2020flash} also quantified 
front-running attacks from an economic point of view,
determining that front-running attacks can also pose a security risk
to the underlying consensus layer by incentivizing unnecessary forks
driven by the maximal extractable value (MEV).

Many previous works explore the idea of 
applying threshold cryptography on \bc.
Virtual ASICs use threshold encryption to implement 
an all-or-nothing broadcast in the blockchain layer~\cite{ganesh2021virtual}.
Sikka~\cite{sikka}, Ferveo (Anoma)~\cite{bebel2022ferveo}, 
Schmid~\cite{noah2021secure}, Dahlia~\cite{malkhi2022maximal}, and Helix~\cite{asayag2018helix} 
apply threshold encryption to mitigate front-running
but only present discussions with specific consensus algorithms.
Fairblock~\cite{momeni2022fairblock} and Shutter~\cite{shutter,shutterL1}
enable encrypted transactions on a per-block basis,
but if an encrypted transaction fails to be included in the sender-chosen block,
then the transaction would be revealed; 
our Strawman III design (\Cref{sec:StrawmanIII}) is based on their approach.

Calypso is a framework 
that enables on-chain secrets 
by adopting threshold encryption 
governed by a \smc~\cite{kokoris2018calypso}. 
Calypso allows ciphertexts to be stored on the blockchain
and collectively decrypted by trustees according to a predefined policy.
\System leverages Calypso to specifically mitigate front-running attacks 
and extends its functionality to release the transaction contents 
once finalized automatically.
\System adopts per-transaction encryption, thus
protecting all unfinalized transactions from front-running attacks,
even if the transactions are delayed.

Other works adopt different approaches 
to mitigate front-running.
A series of recent studies focus on fair 
ordering~\cite{kelkar2021order,kursawe2020wendy,kursawe2021wendy},
but they cannot prevent an adversary 
with a rapid network connection~\cite{baum2021sok}.
Wendy explores the
possibility of combining fair ordering with 
commit-and-reveal~\cite{kursawe2021wendy}
but is in need of quantitative overhead analysis.
Submarine is an application-layer front-running protection approach
that extends a commit-and-reveal design 
to prevent leakage of the smart contract address.
However, it presents a high latency overhead by 
requiring senders to have three rounds of communication with
the \ub~\cite{libsubmarine,breidenbach2018enter}.

Some works adopt time-lock puzzles~\cite{rivest1996time}
to blind transactions.
For example, the injective protocol~\cite{chen2018injective} uses a
verifiable delay function~\cite{boneh2018verifiable}
to achieve a proof-of-elapsed-time.
However, an open challenge remains to link the time-lock puzzle 
parameters to an actual real-world delay~\cite{baum2021sok}.

Finally, works such as MEV-SGX~\cite{mevsgx},
Tesseract~\cite{bentov2019tesseract}, 
Secret Network~\cite{secretnetwork},
and Fairy~\cite{stathakopoulou2021adding} 
use a trusted execution environment~\cite{xing2016intel}
to mitigate front-running.
Nevertheless, these approaches use a centralized component
that is then subject to a single point of failure
or compromise~\cite{ragab2021crosstalk,van2018foreshadow}.
\section{Conclusion}
In this paper, we have introduced \system, a novel blockchain architecture that
addresses front-running attacks with 
TDH2 and PVSS as threshold encryption protocols
on a per-transaction basis.
Our evaluation of \system demonstrates 
that \system is agnostic to consensus algorithms
and to existing smart-contract implementations.
We have also shown that \system meets the necessary
throughput while presenting a low-latency overhead,
thus fitting with Ethereum.
Given that the deployment of \system
would require modifications to a blockchain's execution layer,
\system, in return, would also provide a substantial benefit:
the \system-deployed blockchain would now, by default, 
contain standard front-running
protection for all applications in need at once without requiring any modifications
to smart contracts themselves.

\section*{Acknowledgments}

The authors wish to thank 
Cristina Basescu, Pasindu Nivanthaka Tennage, 
Pierluca Borsò-Tan, and Simone Colombo
for their extremely helpful comments and suggestions
and especially thank Shufan Wang for prototyping \system on the Ethereum blockchain.
This research was supported in part
by U.S. Office of Naval Research grant N00014-19-1-2361,
the AXA Research Fund, 
the PAIDIT project funded by ICRC, 
the IC3-Ethereum Fund,
Algorand Centres of Excellence programme managed by Algorand Foundation,
and armasuisse Science and Technology.
Any opinions, findings, and conclusions or recommendations 
expressed in this material are those of the author(s) 
and do not necessarily reflect 
the views of the funding sources.



\bibliography{main}

\appendix

\newpage

\section{Full Protocol}
\label{sec:fullprotocol}

We provide detailed protocols for \system based on two different threshold cryptographic schemes in this subsection.

\subsection{Protocol based on TDH2}
\label{subsubsection:fullTDHprotocol}
We define a cyclic group $\group$ of prime order $q$ 
with generators $g$ and $\bar{g}$
that are known to all parties, and we
define the following two hash functions:
$\hash_1: \group^5 \times {\left \{0,1\right \}}^l 
\rightarrow \group$ and $\hash_2: \group^3 \rightarrow \ZZ_q$.

\paragraph*{Step 0: DKG Setup}\label{paragraph:dkgsetup}
Before initiating an epoch, 
the \smc runs a DKG protocol to generate a shared public key $\cpk{} = g^{\csk{}}$,
and shares of the private key for each trustee are denoted as $\sk_i$.
The corresponding private key \csk{} can be reconstructed only by combining $t$ 
private key shares.
All trustees also know the verification keys $h_i=g^{\sk_i}$.
We assume that \cpk{} and $h_i$ are written into the blockchain as metadata, 
\eg, in the first block denoting the beginning of this epoch.
We adopt the synchronous DKG protocol 
proposed by Gennaro et al.~\cite{gennaro1999secure}.

\paragraph*{Step 1: Write Transaction}
For the write transaction step, 
we use the encryption protocol presented by the 
TDH2 cryptosystem~\cite{shoup1998securing}.

The sender and the \cg execute the following protocol to write the
\txw on the \ub. The sender then starts the protocol
by performing the following steps to create the transaction \txw:
\begin{enumerate}
    \item Obtain the \smc threshold collective public key \cpk{}
	 from the \ub. 
    \item Generate a symmetric key $k$ and encrypt the signed transaction \tx 
    using authenticated symmetric encryption
    as $c_{\tx} = \enc{k}(\tx)$.
    \item Embed $k$ as a point $k' \in \group$,
    and choose $r,s \in \ZZ_q$ at random.
    \item Compute:
        $$
        c=\cpk{r}k', u=g^r, w=g^s, \bar{u}=\bar{g}^r, 
    	\bar{w}=\bar{g}^s, 
    	$$$$
    	e=\hash_1\left(c,u,\bar{u},w,\bar{w},L\right), f=s+re, 
         $$

        where $L$ is the label of the \ub
        \footnote{This can be the hash of the genesis block.}.
	\item Finally, form the ciphertext $c_k=\left(c,L,u,\bar{u},e,f\right)$
	and construct the write transaction as  $\txw = [c_{\tx}, c_k]_{\sign{\sk_A}}$ 
	signed with the sender's private key ${\sk_A}$.
    \item Send \txw to the \cg.
\end{enumerate}

Upon receiving the \txw, the \cg writes it onto the blockchain.

\paragraph*{Step 2: Shares Preparation by Trustees}

Each trustee $i$ performs the following steps to prepare its decryption share for the \cg.
\begin{enumerate}
    \item Extract $L$ from $c_k$ and verify 
    that $L$ is consistent with the \ub's metadata.
    \item Verify the correctness of the ciphertext $c_k$ using the NIZK
        proof by checking:
        $$
            e=\hash_1\left(c,u,\bar{u},w,\bar{w},L\right) ,
        $$
    	where $w=\frac{g^f}{u^e}$ and 
    	$\bar{w}=\frac{\bar{g}^f}{\bar{u}^e}$.
    	\label{step:KeyReconstructionProof}
	\item If the \txw is valid, choose $s_i \in \ZZ_q$ at 
	random and compute:
 $$
	    u_i=u^{\sk_i}, \hat{u_i}=u^{s_i}, \hat{h_i}=g^{s_i}, 
     $$$$
	    e_i=\hash_2\left(u_i,\hat{u_i},\hat{h_i}\right), f_i=s_i+\sk_i e_i .
     $$

	 \item Create and sign the share:
      ${\mathrm{share}_i} = [u_i, e_i, f_i]_{\sign{\sk_i}}$.
\end{enumerate}

In (\ref{step:KeyReconstructionProof}), the NIZK proof ensures that 
$\log_g u=\log_{\bar{g}} \bar{u}$,
guaranteeing that whoever generated the \txw knows the random value $r$.
If the value of $r$ is known, then the transaction can be decrypted;
as it is impossible to generate \txw without knowing the plaintext transaction,
this property prevents 
replay attacks mentioned in \Cref{sec:ReplayAttack}.

\paragraph*{Step 3: Key Reconstruction}

Once the transaction has received $m$ block confirmations,
each trustee sends their decryption share to the \cg.

Upon receiving the shares, 
each node in the \cg executes the following:
\begin{enumerate}
	\item Each node in the \cg verifies the share by checking:
        $$
            e_i=\hash_2\left(u_i,\hat{u_i},\hat{h_i}\right),
            $$
	where
	$\hat{u_i}=\frac{u^{f_i}}{{u_i}^{e_i}}$ and 
	$\hat{h_i}=\frac{g^{f_i}}{{h_i}^{e_i}}$.
	\label{step:VerifyShare}
	\item After receiving $t$ valid shares, 
	the set of decryption shares is of the form:
     $$
            \{(i,u_i): i \in S\},
            $$
	where $S \subset \{1,...,n\}$ has a cardinality of $t$. 
	Each node then executes the recovery algorithm that does the Lagrange 
	interpolation of the shares:
	$$ \cpk{r}= \prod_{i \in S} {u_i}^{\lambda_i} ,$$
	where $\lambda_i$ is the $i^{th}$ Lagrange element.
	
    \item Recover the encoded encryption key: 
        $$
            k' = c (\cpk{r})^{-1} = (\cpk{r} k') (\cpk{r})^{-1}.
        $$
    \item Retrieve $k$ from $k'$.
\end{enumerate}

In (\ref{step:VerifyShare}), the NIZK proof ensures that 
($u,h_i,u_i$) is a Diffie-Hellman triple, 
\ie, that $u_i=u^{\sk_i}$,
guaranteeing the correctness of the share.

\paragraph*{Step 4: Decryption and Execution}\label{step:execution}
\begin{enumerate}
    \item Decrypt the transaction $tx=\dec{k}(c_{tx})$.
    \item Execute the transaction following the \cg's defined rules.
\end{enumerate}

\subsection{Protocol based on PVSS}\label{fullPVSS}
\label{subsubsection:fullPVSSprotocol}

Let $\mathbb{G}$ be a cyclic group of prime order \textit{q} with two distinct generators \textit{g} and \textit{h}
where the decisional Diffie-Hellman assumption holds.
The \smc has a set of trustees $N = \{1,...,n\}$, where each
trustee is identified by a unique index \textit{i},
and has a private key $sk_i$ and a corresponding public key $pk_i = g^{sk_i}$. 
The \ub stores all the trustees' public keys; 
thus, they are accessible to everyone.
We follow the PVSS scheme presented by Berry Schoenmakers \cite{schoenmakers1999simple}.
The protocol runs as follows:

\paragraph*{Step 0: Share Preparation by Sender}

The sender starts the protocol to prepare key shares and the symmetric key:

\begin{enumerate}
    \item Deriving the generator $h$ from the label of the underlying blockchain $L$ by computing $h = H(L)$ using Elligator maps~\cite{bernstein2013elligator}. 
    This method will protect against replay attacks discussed in \Cref{sec:ReplayAttack}.
    \item Pick a random secret sharing polynomial $s(x) = \sum_{j=0}^{t-1}{a_j x^j}$ of degree at most $t-1$.
    $s = g^{s(0)}$ is the secret to be shared.
    \item Compute the encrypted shares $\hat{s_i} = pk_i^{s(i)}$ of secret $s$ 
    for every secret-management trustee $i$ that the sender wishes to include, 
    create the corresponding NIZK proof $\pi_{\hat{s_i}}$, 
    and the polynomial commitments $b_j = h^{a_j}$, for $0\leq j\leq t-1$.
    \item Use $k$ = H($s$) as the symmetric key.
\end{enumerate}

The NIZK proof $\pi_{\hat{s_i}}$ will be used to verify that the corresponding encrypted share $\hat{s_i}$ is consistent,
\ie, a proof of knowledge of the unique ${s_i}$ that satisfies:
$$
X_i = h^{s(i)} , \hat{s_i} = pk_i^{s(i)}
$$
where $X_i = \prod_{j=0}^{t-1}{b_j}^{i^j} $. $\pi_{\hat{s_i}}$ shows that $\log_h X_i = \log_{pk_i}{\hat{s_i}}$,
and to generate it the sender picks randomly $w_i \in \mathbb{Z}_q$ and computes $a_{1i} = h^{w_i}, a_{2i} = pk_i^{w_i}$. 
Using Fiat-Shamir's technique, the sender then computes the challenge $c_i$ and response $r_i$ as follows:
$$c_i = H(X_i,\hat{s_i}, a_{1i}, a_{2i}), r_i = w_i - s(i)c_i$$
Each proof $\pi_{\hat{s_i}}$ consists of $c_i$ and $r_i$.

\paragraph*{Step 1: Write Transaction}
Once the sender has the $\mathrm{tx}$, 
they can write it to the \ub by the following steps:

\begin{enumerate}
    \item Form the ciphertext $c_k = (\text{<}\hat{s_i}\text{>}, 
    \text{<}\pi_{\hat{s_i}}\text{>}, \text{<}i\text{>}, \text{<}b_j\text{>})$, encrypt the signed transaction tx 
    using authenticated symmetric encryption as $c_\mathrm{tx} = \mathrm{enc}_k(\mathrm{tx})$.
    \item Construct the write transaction as $\mathrm{tx_w} = [c_\mathrm{tx}, c_k]_{\mathrm{sig_{sk}}_A}$ 
    signed with the sender's private key $\mathrm{sk}_A$.
    \item Send \txw to the \cg.
\end{enumerate}

Upon receiving the \txw, the \cg finalizes it onto the blockchain
following its defined consensus rules.

\paragraph*{Step 2: Shares Preparation by Trustees}

Each trustee $i$ performs the following steps to prepare its decryption share.
\begin{enumerate}
    \item Find the corresponding $\hat{s_i}, \pi_{\hat{s_i}}, b_j$ using the index $i$.
    \item Verify the correctness of the encrypted share $\hat{s_i}$ using the NIZK
        proof. Compute $X_i = \prod_{j=0}^{t-1}{b_j}^{i^j}$ from the polynomial commitments 
        $b_j, 0 \leq j < t$. And compute $a_{1i}^{\prime}=h^{r_i} X_i^{c_i}$, $a_{2 i}^{\prime}={pk}_i^{r_i} \hat{s_i}^{c_i}$. 
        Check that H($X_i, \hat{s_i},a_{1i}^{\prime}, a_{2 i}^{\prime}$) matches $c_i$.
    \item If the encrypted share $\hat{s_i}$ is valid, decrypt the share by computing 
         $ s_i = (\hat{s_i})^{{sk_i}^{-1}} $. 
        Create a new NIZK proof $\pi_{s_i}$ to verify the share is correctly decrypted. 
        This proof shows that $\log_g pk_i = \log_{s_i}\hat{s_i}$.
    \item Create and sign the share:
      ${\mathrm{share}_i} = [s_i, \pi_{s_i}]_{\sign{\sk_i}}$.
\end{enumerate}

\paragraph*{Step 3: Key Reconstruction}

Once the transaction has received $m$ block confirmations,
each trustee sends their decryption share to the \cg.

Upon receiving the shares, 
each node in the \cg executes the following:
\begin{enumerate}
	\item  Each node in the \cg verifies the correctness of the decrypted share 
    $\hat{s_i}$ using the NIZK proof $\pi_{s_i}$.
	\item After receiving $t$ valid shares, 
	each node then executes the Lagrange 
	interpolation to recover $s$ from the shares:
	$$ s = \prod_{i = 1}^t {s_i}^{\lambda_i} ,$$
	where $\lambda_i$ is the $i^{th}$ Lagrange element.
	
    \item Recover the encryption key $k =$ H($s$).
\end{enumerate}

\paragraph*{Step 4: Decryption and Execution}
Same as step 4 in \ref{subsubsection:fullTDHprotocol}.

\end{document}